\documentstyle[seceq,preprint]{jpsj}
\input epsf.tex

\makeatletter
\def\lsim{\mathrel{\mathpalette\gl@align<}}
\def\gsim{\mathrel{\mathpalette\gl@align>}}
\def\gl@align#1#2{\lower.6ex\vbox{\baselineskip\z@skip\lineskip\z@
    \ialign{$\m@th#1\hfil##\hfil$\crcr#2\crcr\sim\crcr}}}
\makeatother

\def\runtitle
{Kondo Effect in Single Quantum Dot Systems  --- Study with NRG method ---}

\def\runauthor
{Wataru {\sc Izumida} 
Osamu {\sc Sakai} and Yukihiro {\sc Shimizu}}

\title
{
Kondo Effect in Single Quantum Dot Systems\\
--- Study with Numerical Renormalization Group Method ---}

\author
{Wataru {\sc Izumida}\footnote{izumida@cmpt01.phys.tohoku.ac.jp},
Osamu {\sc Sakai} and Yukihiro {\sc Shimizu}$^{1}$}

\inst
{Department of Physics, Tohoku University, Sendai 980-8578\\
$^1$Department of Applied Physics, Tohoku University, Sendai 980-8579}

\recdate
{March 16, 1998}

\abst
{The tunneling conductance is calculated as a function of
the gate voltage in wide temperature range for the single
quantum dot systems with Coulomb interaction.
We assume that two orbitals are active for the tunneling
process.
We show that the Kondo temperature for each orbital channel
can be largely different.
The tunneling through the Kondo resonance almost fully
develops in the region $T \lsim 0.1 T_{\rm K}^{*} \sim 0.2
T_{\rm K}^{*}$, where $T_{\rm K}^{*}$ is the lowest Kondo
temperature when the gate voltage is varied.
At high temperatures the conductance changes to the usual
Coulomb oscillations type.
In the intermediate temperature region, the degree of the
coherency of each orbital channel is different, so strange
behaviors of the conductance can appear.
For example, the conductance once increases and then
decreases with temperature decreasing when it is suppressed
at $T=0$ by the interference cancellation between different
channels.
The interaction effects in the quantum dot systems lead the
sensitivities of the conductance to the temperature and to
the gate voltage.}

\kword
{quantum dot, tunneling, Kondo effect, Coulomb oscillations}

\begin{document}
\sloppy
\maketitle

\section{Introduction}
\label{sec:Intro}

Coulomb oscillations behavior in quantum dot systems is well
known as one of the typical phenomena due to the Coulomb
repulsion between electrons in the dot.
But it is one aspect of the electron-electron interaction
effects.
Local spin moment is induced by the Coulomb repulsion in the
dot sometime.
It will fluctuate by exchanging spin moment with leads and
thus cause the strong inelastic scattering.
But the local spin disappears in the lowest temperature due
to the spin singlet formation between leads and the dot.
These effects have been investigated as the Kondo effect for
the magnetic impurity systems in many
years,~\cite{rf:Rev.Kondo_Effect} and for the quantum dot
systems
recently.~\cite{rf:Ng,rf:Glazman,rf:Kawabata,rf:Hershfield,rf:Yeyati,rf:Meir_Wingreen,rf:Inoshita,rf:Oguri.1,rf:Izumida,rf:Kondo-Kastner}

The origins of the non-vanishing spin moment in the dot are
not unique.
Of course unpaired electron in the odd number electrons in
the dot brings the spin moment.
Furthermore, degenerate orbitals, or strong Hund's rule
coupling can induce the non-vanishing spin moment even in
the even electron number cases.~\cite{rf:Tarucha}
This will cause varieties in the behavior of the spin state
when the electron number is changed by the gate voltage.
In the low temperature range that the Kondo singlet state is
formed, the inelastic scattering due to the spin fluctuation
is suppressed.
Therefore the coherent process will dominate the tunneling.
This process is considered as the resonant tunneling through
the Kondo peak at the Fermi level in some sense.
The Kondo peak sensitively diminishes as the temperature
increases and disappears at high temperatures because the
characteristic temperature for the many body singlet
formation, the Kondo temperature, is usually very low.
In general, the tunneling conductance will show various
types of the interference effect depending on the geometry
and the connectivity of the dot systems.
The coherency of the system is disturbed by the inelastic
scattering due to the spin moment revived at finite
temperatures.
However, calculation of the conductance in such the
intermediate temperature has not been done theoretically
because the reliable method to treat the Kondo effect has
not been established.

\begin{figure}[htb]
   \centerline{\epsfxsize=3.25in\epsfbox{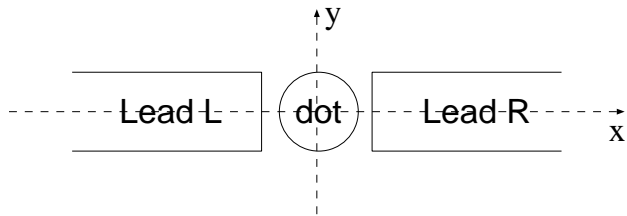}}
   \caption
   {
   The configuration of the system.
   The $z$-axis is perpendicular to the $xy$ plane.
   }
   \label{fig:system}
\end{figure}

In this paper, we consider the system shown in Fig. 
{\ref{fig:system}}, a quantum dot and connected two leads.
We calculate the tunneling conductance in wide temperature
range from the high temperature region in which the usual
Coulomb oscillations are observed, to the lower temperature
region in which the Kondo singlet state is formed.
Conductance is calculated as the function of the gate
voltage.
We assume that two orbitals of the dot are active for the
tunneling process.
The following two cases are considered for the tunneling
Hamiltonian (TH) between the conduction channels in the
leads and the orbitals in the dot:
(THi) There is one conduction channel in each lead, and it
hybridizes with both of two orbitals in the dot.
This means that the tunneling processes {\it via} the two
orbitals interfere each other.
(THii) There are two conduction channels in each lead, and
each of the channels hybridizes separately with one of two
orbitals in the dot.
In this case the tunneling processes of the two channels are
independent and do not interfere.
(But the electronic state of the dot itself is affected by
the interaction between electrons in the different
channels.)
In such two situations, we study the following three cases
for the orbital state in the dot; (1) orbital energies in
the dot are not degenerate, (2) orbital energies in the dot
are degenerate, (3) electrons on the orbitals have strong
Hund's rule coupling.
The six combinations of the situation, from THi-1 to THii-3,
are studied systematically.

There have been several methods to calculate the conductance
in finite temperatures; perturbation theory, non-crossing
approximation, quantum Monte Calro method, numerical
renormalization group (NRG)
method.~\cite{rf:Hershfield,rf:Yeyati,rf:Meir_Wingreen,rf:Inoshita,rf:Oguri.1,rf:Izumida}
The NRG is a reliable method in the temperature region $T
\lsim T_{\rm K}$, where $T_{\rm K}$ is the Kondo
temperature.~\cite{rf:Izumida,rf:NRG_Kri,rf:NRG_sakai,rf:NRG_shimizu}
This method directly calculates the tunneling matrices
between wave functions of many electron states.
Therefore we can obtain the tunneling probabilities in a
general way for the coherent process in the low temperatures
and the incoherent process in the high temperatures.
The numerical calculations are performed by using NRG method
through this paper.

We found that the tunneling through the Kondo resonance
fully develops in the temperature region, $T \lsim 0.1
T_{\rm K}^{*} \sim 0.2 T_{\rm K}^{*}$, where $T_{\rm K}^{*}$
is the lowest Kondo temperature of the dot system when the
gate voltage is varied.
The interference effect of the tunneling process also
becomes important in this region.
In high temperature region $T \gg T_{\rm K}^{*}$, the
Coulomb oscillations type conductance is observed commonly.
In the intermediate temperature range among these, various
types of conductance behaviors appear, especially in THi
cases, because the coherency of each orbital channel is
sensitively changed by the temperature and the gate voltage.

In \S \ref{sec:One_channel}, the cases classified to the THi
are studied, and the cases classified to the THii are
studied in \S \ref{sec:Two_channel}.
In \S \ref{sec:Summary}, summary and discussion of this
study are given.

\section{One Conduction Channel in Each Lead}
\label{sec:One_channel}

To study the system shown in Fig. {\ref{fig:system}},
following things are contained in the model:
i) Electronic states in the two leads are written as the
conduction band states.  ii) There are two orbitals in the
dot.  iii) Electron-electron interactions in the dot are
considered.  And, iv) the electrons have tunneling matrices
between the leads and the dot.
For simplicity, we consider the situation that there are
only two orbitals in the dot instead of considering many
orbitals.
We assume that the system has the mirror symmetry with
respect to the $yz$ and $zx$ planes.
(The $z$-axis is perpendicular to the $xy$ plane.)

In this section, we discuss the situation classified to the
THi that there are only one conduction channel in each lead,
and tunneling processes {\it via} the different orbital
channels interfere each other.
We consider the one-dimensional degree of freedom in each
lead.
This situation would appear in the case that leads are very
narrow, or the joint part between the leads and the dot are
very narrow.

\subsection{Model Hamiltonian}
\label{sec:model_H}

We consider the two orbitals in the dot, which are even and
odd, respectively, under the $yz$ mirror, and are even under
the $zx$ mirror.
These orbitals in the dot are named as the even orbital and
the odd orbital, respectively.
We consider the one-dimensional degree of freedom with even
symmetry under the $zx$ mirror in each lead.
We use the following Anderson Hamiltonian,
\begin{eqnarray}
   H & = & H_{\rm l} + H_{\rm d} + H_{\rm l-d},
   \label{eq:total_H}\\
   H_{\rm l} & = & 
                     \sum_{k \sigma} 
                     \varepsilon_{k} 
                     c_{{\rm L} k \sigma}^{\dagger} c_{{\rm L} k \sigma}
                 +
                     \sum_{k \sigma} 
                     \varepsilon_{k} 
                     c_{{\rm R} k \sigma}^{\dagger} c_{{\rm R} k \sigma},
   \label{eq:H_l_2cnl}\\
   H_{\rm d} & = & 
                     (\varepsilon_{\rm d} - t) 
                     \sum_{\sigma}
                     n_{{\rm e} \sigma}
                 +
                     (\varepsilon_{\rm d} + t) 
                     \sum_{\sigma}
                     n_{{\rm o} \sigma}
   \nonumber\\
              && +
                     U
                     \sum_{p={\rm e,o}}
                     n_{p \uparrow} n_{p \downarrow}
                 +
                     U
                     \sum_{ \sigma, \sigma^{'} } 
                     n_{{\rm e} \sigma} n_{{\rm o} \sigma^{'}}
   \nonumber\\
              && +
                     J_{\rm H} 
                     \sum_{\sigma_{1}\sigma_{2}\sigma_{3}\sigma_{4}}
                     ({\mbox{\boldmath $\sigma$}})_{\sigma_{1}\sigma_{2}} 
                     \cdot ({\mbox{\boldmath $\sigma$}})_{\sigma_{3}\sigma_{4}}
                     d_{{\rm e} \sigma_{1}}^{\dagger} d_{{\rm e} \sigma_{2}} 
                     d_{{\rm o} \sigma_{3}}^{\dagger} d_{{\rm o} \sigma_{4}},
   \label{eq:H_d_2cnl}\\
   H_{\rm l-d} & = &
                       \frac{1}{\sqrt{2}}
                       \sum_{k \sigma}
                            V_{\rm e} d_{{\rm e} \sigma}^{\dagger} 
                            \left( c_{{\rm L} k \sigma} + c_{{\rm R} k \sigma} \right)
                   +
                       \frac{1}{\sqrt{2}}
                       \sum_{k \sigma}
                            V_{\rm o} d_{{\rm o} \sigma}^{\dagger} 
                            \left( c_{{\rm L} k \sigma} - c_{{\rm R} k \sigma} \right)
                   + {\rm h.c.}
   \label{eq:H_l-d_2cnl}
\end{eqnarray}
The terms $H_{\rm l}$ and $H_{\rm d}$ give the electron
orbitals in the two leads and in the dot, respectively.
The term $H_{{\rm l-d}}$ gives the electron tunneling
between the two leads and the dot.
The annihilation operator of the left (right) lead state is
denoted by $c_{{\rm L} k \sigma}$ ($c_{{\rm R} k \sigma}$),
and that of the even (odd) orbital in the dot is denoted by
$d_{{\rm e} \sigma}$ ($d_{{\rm o} \sigma}$).
The quantity $\varepsilon_{k}$ is the energy of the lead
state.
The quantity $\varepsilon_{\rm d}$ corresponds to dot's
potential, and is able to change by applying gate voltage.
The energy level of the even (odd) orbital is given by
$\varepsilon_{\rm d}-t$ ($\varepsilon_{\rm d}+t$).
The energy separation between the even and odd orbitals is
defined as $\Delta \varepsilon_{\rm d} \equiv 2 t$.
The Coulomb interaction constant is given by $U$ for both
the intra- and inter-orbital terms.
We consider the exchange interaction between electrons on
the even and odd orbitals with strength $J_{\rm H}$.
The operator ${\mbox{\boldmath $\sigma$}}$ is the Pauli
matrix.
The quantity $V_{\rm e}$ ($V_{\rm o}$) is the hybridization
matrix for the even (odd) orbital.
(We neglect the $k$ dependence of the tunneling matrices.)

We introduce the even and the odd combinations of the lead
orbitals, $s_{k \sigma} = (c_{{\rm L} k \sigma} + c_{{\rm R}
k \sigma}) / \sqrt{2}$ and $a_{k \sigma} = (c_{{\rm L} k
\sigma} - c_{{\rm R} k \sigma}) / \sqrt{2}$, respectively.
The terms $H_{\rm l}$ and $H_{\rm l-d}$ are rewritten as,
\begin{eqnarray}
   H_{{\rm l}} & = & 
                     \sum_{k \sigma}
                     \varepsilon_{k}
                     s_{k \sigma}^{\dagger} s_{k \sigma}
                 +
                     \sum_{k \sigma}
                     \varepsilon_{k}
                     a_{k \sigma}^{\dagger} a_{k \sigma},
   \label{eq:H_l_1}\\
   H_{{\rm l-d}} & = & 
                       \sum_{k \sigma}
                       \left(
                       V_{\rm e} d_{{\rm e} \sigma}^{\dagger} s_{k \sigma} + {\rm h.c.}
                             \right)
                     +
                       \sum_{k \sigma}
                       \left(
                       V_{\rm o} d_{{\rm o} \sigma}^{\dagger} a_{k \sigma} + {\rm h.c.}
                             \right).
   \label{eq:H_l-d_1}
\end{eqnarray}
The hybridization strength for the even (odd) orbital
channel is denoted as $\Delta_{\rm e} = \pi V_{\rm e}^2
\rho(\varepsilon_{\rm F})$ ($\Delta_{\rm o} = \pi V_{\rm
o}^2 \rho(\varepsilon_{\rm F})$), where
$\rho(\varepsilon_{\rm F})$ is the density of states of the
lead states.
Hereafter we denote the even and the odd orbitals by $p$ as
$p = {\rm e}$ and $p = {\rm o}$, respectively.

\subsection{Linear response conductance}
\label{sec:conductance_F}

In this paper we restrict ourselves to the linear response
conductance for the applied bias voltage between the left
and the right leads.
We generally define an electric current from the left lead L
to the right lead R as follows;
\begin{eqnarray}
   I 
   & \equiv & 
   -e \frac{ -\langle \dot{N_{\rm L}} \rangle
   +\langle \dot{N_{\rm R}} \rangle }{2}, \label{eq:def;current}
\end{eqnarray}
where $-e$ is the charge of an electron, $\langle
\dot{N_{\rm L}} \rangle$ is the expectation value of the
time differentiation of the electron number operator in the
lead L.
The quantity $\langle \dot{N_{\rm R}} \rangle$ is that of
the lead R.
We obtain the following expression for the linear response
conductance formula,~\cite{rf:Kubo}
\begin{eqnarray}
   G & \equiv & \frac{I}{2V} \nonumber \\
       & = &    \frac{2 e^{2}}{h} 
                \lim_{\omega \rightarrow 0} 
                \frac{P^{''}(\omega)}{\omega}, \label{eq:G}
\end{eqnarray}
with
\begin{eqnarray}
   P^{''}(\omega) & = & 
                     \frac{\pi^{2} \hbar^{2}}{4}
                     \frac{1}{Z} \sum_{n,m} 
                     \left( {\rm e}^{- \beta E_{m}} - {\rm e}^{- \beta E_{n}} \right) 
   \nonumber \\
               && 
                     \times 
                     \left| \langle n \left| 
                     \dot{N_{\rm L}} - \dot{N_{\rm R}} 
                     \right| m \rangle \right| ^{2} 
   \nonumber \\
               &&    \times 
                     \delta \left( \omega - \left( E_{n} - E_{m} \right) \right),
   \label{eq:P''}
\end{eqnarray}
where $Z = \sum_{n} e^{- \beta E_{n}}$ is the partition
function, $\beta$ is the inverse of the temperature,
$\beta=1/T$.
(See Appendix of the ref. \citen{rf:Izumida} for the
derivation of eqs. (\ref{eq:G}) and (\ref{eq:P''}).)
We note that we do not have done restrictive assumptions for
the electric current such as the `coherent tunneling' or the
`sequential tunneling' in the derivation of (\ref{eq:G}).
The conductance for the case of THi, $G^{(1)}$, is
calculated by using eqs. (\ref{eq:G}) and (\ref{eq:P''})
with $N_{\rm L}=\sum_{k \sigma} c_{{\rm L} k
\sigma}^{\dagger} c_{{\rm L} k \sigma}$ and $N_{\rm
R}=\sum_{k \sigma} c_{{\rm R} k \sigma}^{\dagger} c_{{\rm R}
k \sigma}$.
We note that the tunneling processes {\it via} the even and
the odd orbitals interfere each other in the tunneling
between leads.
Following the method described in ref. \citen{rf:Izumida},
the current spectrum, $P^{''}(\omega)$, is calculated by NRG
method.

Here we briefly consider the tunneling at absolute zero
temperature.
The system would be the local Fermi liquid state at very low
temperatures.
In this case there are no spin scattering processes.
In such the coherent tunneling process at zero temperature,
the conductance for the case of THi, $G_{\rm F}^{(1)}$, is
written as follows;~\cite{rf:Kawabata,rf:Izumida}
\begin{eqnarray}
 G_{\rm F}^{(1)}
           & = &
                 \frac{2 e^{2}}{h}
                 \sin^{2}
                 \left\{
                         \frac{\pi}{2}
                         \left( 
                                 \langle n_{\rm e} \rangle 
                               - \langle n_{\rm o} \rangle 
                         \right)
                 \right\}. \label{eq:G_F}
\end{eqnarray}
This expression is the extension of the relation $G_{\rm F}
= (2 e^{2} / h) \sin^{2} \left\{ \left( \pi/2 \right)
\langle n \rangle \right\}$ for the single orbital
case.~\cite{rf:Ng,rf:Glazman}
The quantity $\langle n_{\rm e} \rangle - \langle n_{\rm o}
\rangle$ appears through the interference of the tunneling
processes {\it via} the even and odd orbitals.
In derivation of eq. (\ref{eq:G_F}), we have applied
Friedel's sum rule to the even and the odd orbital channels
separately.~\cite{rf:Friedel_Lanhreth,rf:Friedel_Shiba,rf:2chlFriedel}

\subsection{Numerical results with NRG method}
\label{sec:Result_1}

\subsubsection{Non-degenerate orbital case}
\label{sec:result_1_1}

In this subsection we present the numerical results for the
non-degenerate orbital case.
The parameters are chosen to satisfy the relations
$\Delta_{\rm e}, \Delta_{\rm o} \ll \Delta\varepsilon_{\rm
d}$: $\Delta_{\rm e} = 0.003 \pi$, $\Delta_{\rm o} = 0.002
\pi$, $\Delta\varepsilon_{\rm d} = 0.1$, $U = 0.1$ and
$J_{\rm H} = 0$.
(In this paper we choose the band width $D$ as an energy
unit.
The density of states for the conduction bands in each
orbital channel are assumed to be constant within the region
$ -1 < \omega / D < 1$.)

\begin{figure}[htb]
   \centerline{\epsfxsize=3.25in\epsfbox{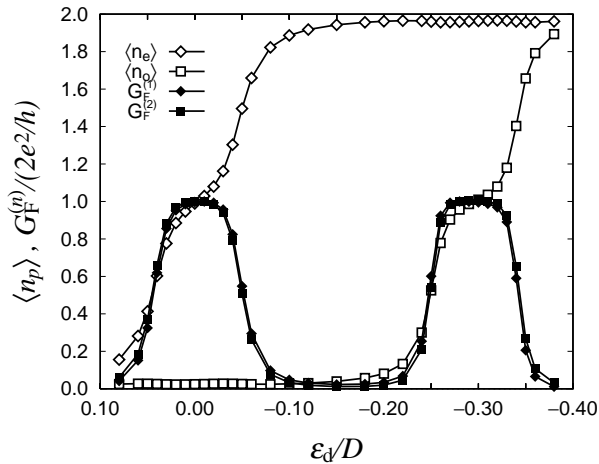}}
   \caption
   {
    The expectation value of the electron occupation numbers $\langle n_{p} \rangle$
    for the even ($p={\rm e}$) and the odd ($p={\rm o}$) orbitals,
    and the conductance $G_{\rm F}^{(n)}$
    for the THi case ($n=1$) and the THii case ($n=2$),
    as a function of 
    the potential $\varepsilon_{\rm d}$
    at $T=0$ for the non-degenerate orbital case.
   }
   \label{fig:exn_2lv_nod}
\end{figure}

First we present the electron occupation numbers at absolute
zero temperature, $\langle n_{p} \rangle$, as a function of
the dot's potential, $\varepsilon_{\rm d}$.
The conductance $G_{\rm F}^{(1)}$ at $T=0$, which is
calculated from (\ref{eq:G_F}), is also shown in Fig. 
{\ref{fig:exn_2lv_nod}}.
(The conductance $G_{\rm F}^{(2)}$ at $T=0$ for the case of
THii is also shown in the same figure.
This quantity will be discussed in \S
\ref{sec:Two_channel}.)
As the potential $\varepsilon_{\rm d}$ is gradually
decreased by applying gate voltage, the first electron
occupies the even orbital at about $\varepsilon_{\rm d} \sim
t = 0.05$ ($\langle n_{\rm e} \rangle \sim 0.5$), and then
the second electron occupies at about $\varepsilon_{\rm d}
\sim t - U = -0.05$ ($\langle n_{\rm e} \rangle \sim 1.5$).
The third and the forth electrons occupy the odd orbital at
about $\varepsilon_{\rm d} \sim - t - 2U = -0.25$ ($\langle
n_{\rm o} \rangle \sim 0.5$) and at $\varepsilon_{\rm d} =
-t - 3U = -0.35$ ($\langle n_{\rm o} \rangle \sim 1.5$),
respectively.
(The potential values $\varepsilon_{\rm d} = t, t-U, -t-2U$
and $-t-3U$ are just the cross points of energies of the
states with different electron number in the dot, if one
neglects the hybridizations between leads and dot.)
The conductance $G_{\rm F}^{(1)}$ has the finite value $\sim
2e^{2} / h$ (unitarity limit value) in the region of $t - U
\lsim \varepsilon_{\rm d} \lsim t$ in which the even orbital
is almost half-filled, and in the region of $-t - 3U \lsim
\varepsilon_{\rm d} \lsim -t - 2U$ in which the odd orbital
is almost half-filled.
Both cases have odd electron numbers.
Therefore, there will appear spin moment when the
temperature rises near the Kondo temperature.
This will cause inelastic scattering process and will
strikingly change the conductance from that of $T=0$.

\begin{figure}[htb]
   \centerline{\epsfxsize=3.25in\epsfbox{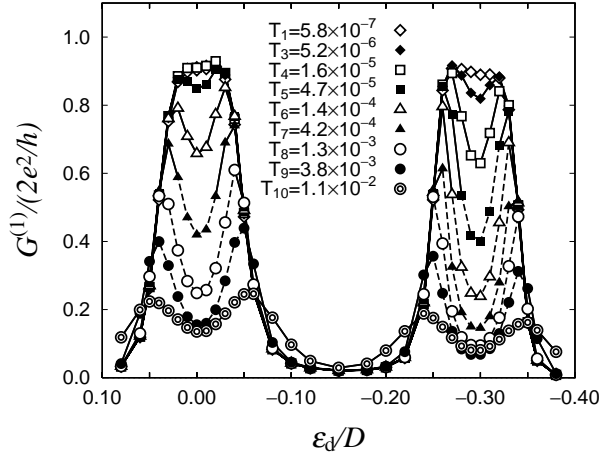}}
   \caption
   {
    The temperature dependence of the conductance
    as a function of the parameter $\varepsilon_{\rm d}$
    for the THi-1 case.
    (For the non-degenerate orbital case.)
    The regions given by broken lines are considered to be the magnetic state 
    and solid lines region are the non-magnetic state (spin singlet state).
    This classification is done by comparing the temperature 
    and the Kondo temperature estimated
    from the magnetic excitation spectra.
    The suffixes for the temperature are chosen to be common 
    throughout \S \ref{sec:result_1_1} 
    and Fig. \ref{fig:cond_2ch_nod} in \S \ref{sec:Two_channel}.
   }
   \label{fig:cond_2lv_nod}
\end{figure}

Next, we present the temperature dependence of the
conductance in Fig. {\ref{fig:cond_2lv_nod}}.
It is calculated from the current spectrum $P^{''}(\omega)$
by using eqs. ({\ref{eq:G}}) and ({\ref{eq:P''}}).
Let us see Fig. {\ref{fig:cond_2lv_nod}} from high
temperature side.
There are four peaks at the potentials $\varepsilon_{\rm d}
\sim 0.05, -0.05, -0.25$ and $-0.35$ in high temperature
region $T \gsim 10^{-3}$.
These structures correspond to the usual Coulomb
oscillations, because the each potential coincides with the
point at which the electron number in the dot changes.
The thermally broadened peak structures in $T_{10}=1.1
\times 10^{-2}$ case change to more sharp peak structures in
$T_{9}=3.8 \times 10^{-3}$ case.
When the temperature gradually decreases below about $T \sim
10^{-3}$, the conductance grows up in the regions $t - U
\lsim \varepsilon_{\rm d} \lsim t$ and $-t - 3U \lsim
\varepsilon_{\rm d} \lsim -t - 2U$.
The pairs of the peaks at high temperatures in the each
region gradually change to the one peak structures.
In the lowest temperature limit $T \lsim 10^{-6}$, the
conductance $G^{(1)}$ coincides to $G_{\rm F}^{(1)}$ in Fig. 
{\ref{fig:exn_2lv_nod}} except the numerical errors brought
by the calculation of $P^{''}(\omega) / \omega$ with NRG
method.
(From comparison of the conductance $G_{\rm F}^{(1)}$ in
Fig. {\ref{fig:exn_2lv_nod}} and $G^{(1)}$ at $T_{1}=5.8
\times 10^{-7}$ in Fig. {\ref{fig:cond_2lv_nod}}, $G^{(1)}$
has about $15{\%}$ smaller magnitude at $\varepsilon_{\rm
d}=-0.30$.
This accuracy of the conductance through the calculation of
$P^{''}(\omega) / \omega$ is not so good in a quantitative
sense, but seems to be sufficient to extract the qualitative
temperature dependence caused by the change of the
electronic state.)
As shown later, the Kondo temperature $T_{\rm K}$ strongly
depends on $\varepsilon_{\rm d}$.
Hereafter, we denote $T_{\rm K}^{*}$ as the lowest Kondo
temperature when the potential $\varepsilon_{\rm d}$ is
changed.
It is estimated to be about $T_{\rm K}^{*} = 2.7 \times
10^{-5}$ for $\varepsilon_{\rm d}=-0.30$ case, which is near
$T_{4}=1.6 \times 10^{-5}$ in Fig. {\ref{fig:cond_2lv_nod}}.
The valley of the conductance at $\varepsilon_{\rm d}=-0.30$
becomes very shallow in the cases $T \lsim T_{3} \sim 0.2
T_{\rm K}^{*}$.
The conductance changes from the usual Coulomb oscillations
type to the Kondo resonance type at about $T \sim 0.2 T_{\rm
K}^{*}$.
In the non-degenerate orbital case of this subsection, the
interference between tunneling processes {\it via} the even
and odd orbital channels seems to be not so operative.
We expect that the temperature dependence of the growing up
of the pair peaks to the flat one peak is almost identical
to that of the single orbital case.

Next we present the excitation spectra in detail.
Hereafter in this subsection, we mainly discuss only the
region $\varepsilon_{\rm d} \le -3U/2 = -0.15$.
In this case the Kondo effect of the odd channel electron is
important.
We have calculated the single particle excitation spectra
for the even and the odd orbitals, $\rho_{p}(\omega)$
($p={\rm e ,o}$), which are given as follows,
\begin{eqnarray}
    \rho_{p}(\omega)
      & \equiv &
    \frac{1}{Z} \sum_{n,m}
    \left( {\rm e}^{-\beta E_{m}} - {\rm e}^{-\beta E_{n}} \right) \nonumber\\
      &&
    \times
    \left( \left| \langle n | d_{p}^{\dagger} | m \rangle \right|^{2}
           \delta \left( \omega- \left( E_{n}-E_{m} \right) \right) \right. \nonumber \\
      &&  
    \left. +
    \left| \langle n | d_{p} | m \rangle \right|^{2}
    \delta \left( \omega+ \left( E_{n}-E_{m} \right) \right)
    \right). \label{eq:rho}
\end{eqnarray}
The single particle excitation spectra at several
temperatures for the case of $\varepsilon_{\rm d} = -0.30$
are shown in Fig. {\ref{fig:spctr_Kondo.nod}}.
\begin{fullfigure}[htb]
   \centerline{\epsfxsize=7in\epsfbox{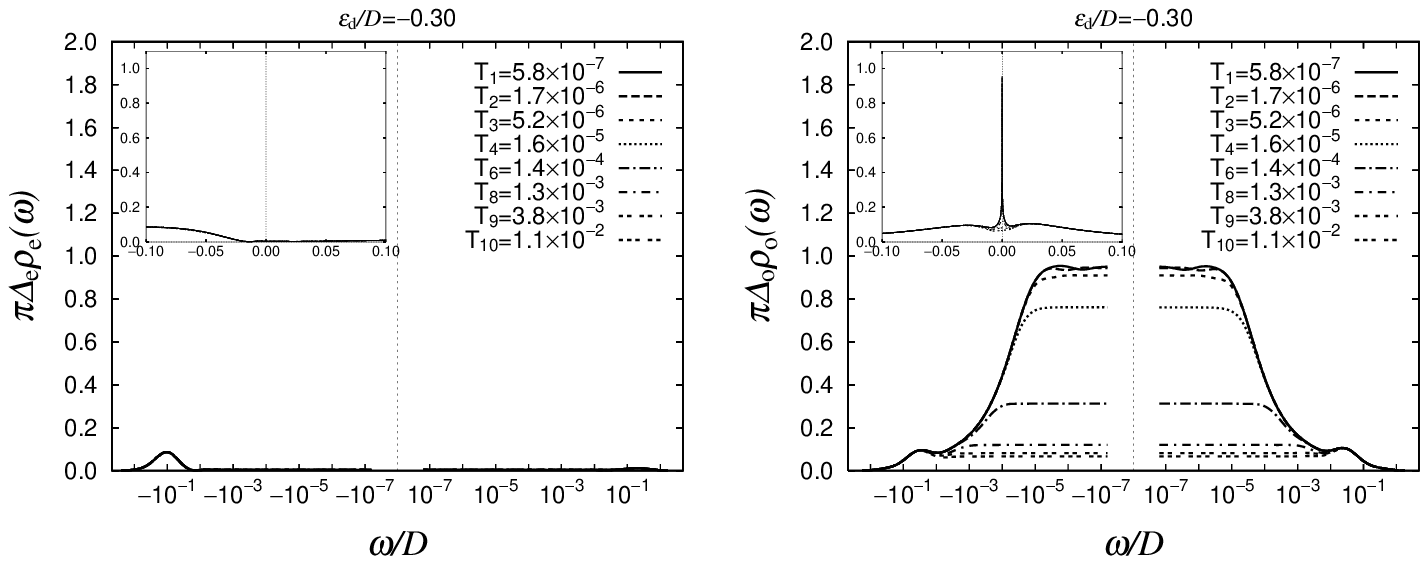}}
   \caption
   {
    The single particle excitation spectra at several temperatures 
    for the potential $\varepsilon_{\rm d} = -0.30$
    ($\langle n_{\rm o} \rangle \sim 1.0$,
    in which the odd channel is the Kondo regime).
    The results are for the non-degenerate orbital case.
    The spectra are plotted as a function of the logarithm of the energy
    except the inset plotted by the energy.
   }
   \label{fig:spctr_Kondo.nod}
\end{fullfigure}
In $\rho_{\rm e}(\omega)$ we have no fine structures near
the Fermi energy because the even orbital is almost
full-filled.
On the other hand the odd orbital is almost half-filled,
$\langle n_{\rm o} \rangle \sim 1.0$, $\rho_{\rm o}(\omega)$
has fine structures at low temperatures.
The Kondo peak on the Fermi energy gradually grows up in the
energy region $|\omega| \lsim 10^{-4}$ as temperature
decreases from $T \sim 10^{-4}$.
Growing up of the Kondo peak completes at about $T \lsim
10^{-6}$.
Reflecting these facts, the conductance near
$\varepsilon_{\rm d} \sim -0.30$ increases as temperature
decreases.
The odd channel seems to be in the Kondo regime.
In Fig. {\ref{fig:spctr_Mix.nod} the potential is raised to
$\varepsilon_{\rm d} = -0.26$ so that the odd channel is in
the mixed valence regime ($\langle n_{\rm o} \rangle \sim
0.8$).
\begin{fullfigure}[htb]
   \centerline{\epsfxsize=7.00in\epsfbox{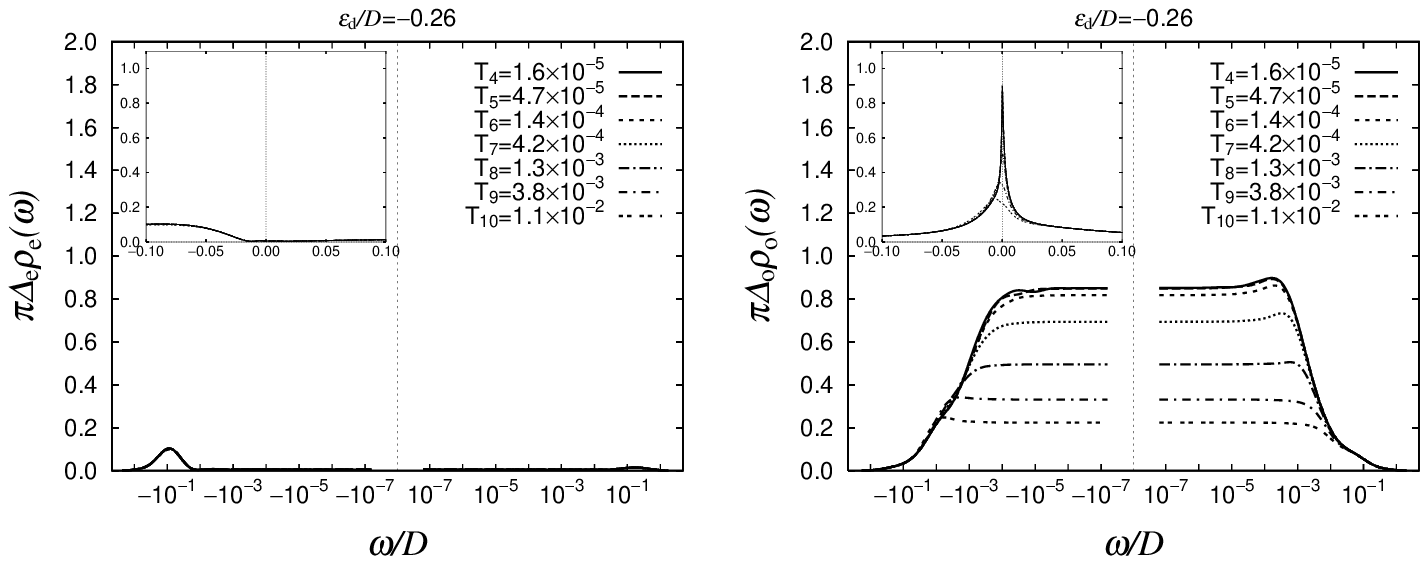}}
   \caption
   {
    The single particle excitation spectra at several temperatures 
    for the potential $\varepsilon_{\rm d} = -0.26$
    ($\langle n_{\rm o} \rangle \sim 0.8$).
    The results are for the non-degenerate orbital case.
    In this case the odd channel is in the mixed valence regime.
   }
   \label{fig:spctr_Mix.nod}
\end{fullfigure}
In $\rho_{\rm o}(\omega)$, we can see an excitation peak
near $\omega \sim -0.01$ in the case of $T_{10} = 1.1 \times
10^{-2}$.
This corresponds to the excitation from the $N_{\rm o}=1$
state to the $N_{\rm o}=0$ state.
As the temperature decreases gradually, the peak position
shifts to the Fermi energy, and the peak height increases.
It grows up completely in the temperature region $T \lsim
10^{-4}$, and has structures seem to be a composite of the
two components: the electron excitation of $N_{\rm o}=1
\rightarrow N_{\rm o}=0$ and the Kondo peak.
As seen in Fig. {\ref{fig:spctr_Kondo.nod}}, these two
components are separated with each other when the system is
in the Kondo regime.
Any way, the intensity of $\rho_{\rm o}(\omega)$ on the
Fermi energy increases as temperature decreases, and causes
the growth of the conductance in the potential region $-t-3U
\lsim \varepsilon_{\rm d} \lsim -t-2U$.

As seen from comparison of Fig. {\ref{fig:spctr_Kondo.nod}}
and Fig. {\ref{fig:spctr_Mix.nod}}, the Kondo temperature is
different in each potential case.
To see this point clearly we calculate the magnetic
excitation spectrum at zero temperature, $\chi_{\rm
m}^{''}(\omega)$,
\begin{eqnarray}
   \chi_{\rm m}^{''}(\omega)
      & \equiv &
   \sum_{n} \sum_{{\rm Gr}}
   \left| \langle n \left|
          \left( S_{{\rm e},z} + S_{{\rm o},z} \right)
                    \right| {\rm Gr} \rangle 
   \right|^{2} \nonumber \\
      && 
   \times \delta \left( \omega- \left( E_{n}-E_{{\rm Gr}} \right) \right),
   \label{eq:chi}
\end{eqnarray}
where $S_{p,z} = (d_{p,\uparrow}^{\dagger} d_{p,\uparrow} -
d_{p,\downarrow}^{\dagger} d_{p,\downarrow})/2$ is the spin
operator on the $p$-th orbital and ${\rm Gr}$ denotes the
ground state of the system.
The peak position of the magnetic excitation reflects the
characteristic energy of the spin fluctuation.
The Kondo temperature $T_{\rm K}$ is defined as the energy
of the peak position.~\cite{rf:Izumida,rf:Kondo_Temp.}
It is shown in Fig. {\ref{fig:tk_nod}} as a function of the
potential $\varepsilon_{\rm d}$, and has the sharp local
minima at $\varepsilon_{\rm d} = -0.30$ (where $\langle
n_{\rm o} \rangle \sim 1.0$) and $\varepsilon_{\rm d} =
0.00$ ($\langle n_{\rm e} \rangle \sim 1.0$).
\begin{figure}[htb]
   \centerline{\epsfxsize=3.25in\epsfbox{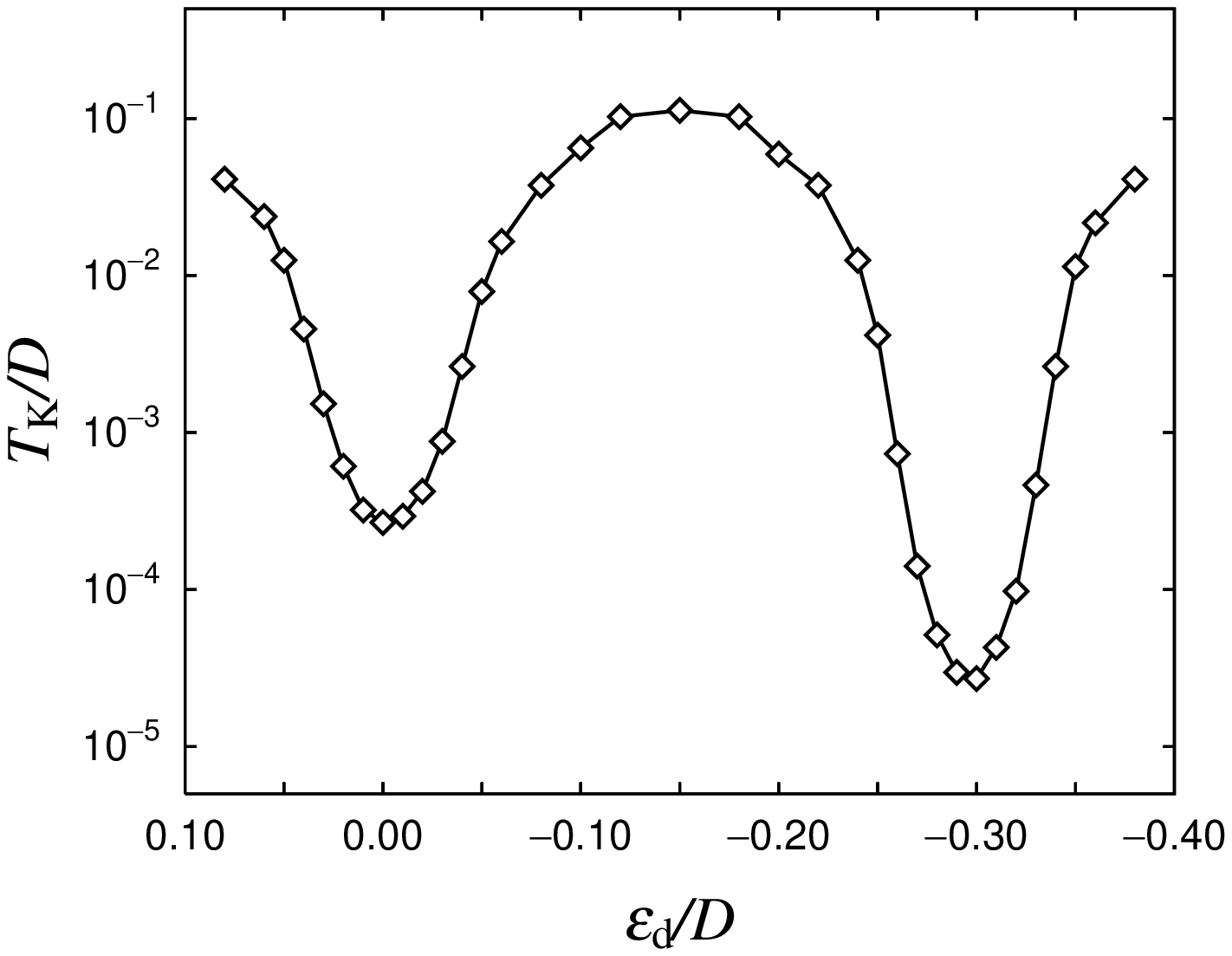}}
   \caption
   {
    The Kondo temperature as a function 
    of the potential $\varepsilon_{\rm d}$
    for the non-degenerate orbital case.
   }
   \label{fig:tk_nod}
\end{figure}
These minima characterize the temperature at which the
conductance behavior changes from the Coulomb oscillations
type to the Kondo resonance type.

We point out that the Kondo temperature is also very
sensitive to the change of the hybridization $\Delta$ as
seen from comparison of the $T_{\rm K}(\varepsilon_{\rm
d}=-0.30)=2.7 \times 10^{-5}$ ($\Delta_{\rm o}=0.002 \pi$)
and $T_{\rm K}(\varepsilon_{\rm d}= 0.00)=2.7 \times
10^{-4}$ ($\Delta_{\rm e}=0.003 \pi$), or from the
analytical expression for the Kondo temperature.
($T_{\rm K} = \sqrt{U \Delta / 2} \exp{ [ -(\pi U)/(8\Delta)
+ \Delta/(2 U) ]}$ for the single orbital impurity Anderson
model in the electron-hole symmetric
case~\cite{rf:kawakami}).
This means that the Kondo temperature of the quantum dot
systems is sensitive to the effective distance between leads
and dot.

\subsubsection{Degenerate orbital case}
\label{sec:result_1_2}

Next, we consider the situation that the two orbitals have
degenerate energy levels.
Even if the orbital energies are not degenerate in the
strict sense, they should be considered as to be degenerate
when the energy difference $\Delta \varepsilon_{\rm d}$ is
less than $T_{\rm K}$.
(See in Appendix.)
The effect of spin fluctuation will appear even when
electron number in the dot is even.
In this subsection, we present the numerical results for the
degenerate orbital case.
The parameters are chosen as $\Delta_{\rm e} = 0.003 \pi$,
$\Delta_{\rm o} = 0.002 \pi$, $\Delta\varepsilon_{\rm d}
(=2t) = 0$, $U = 0.1$ and $J_{\rm H} = 0$ for numerical
calculation.

The electron occupation numbers for the even and the odd
orbitals increase almost simultaneously when the potential
$\varepsilon_{\rm d}$ is dropped as shown in Fig. 
{\ref{fig:exn_2lv_od}}.
\begin{figure}[htb]
   \centerline{\epsfxsize=3.25in\epsfbox{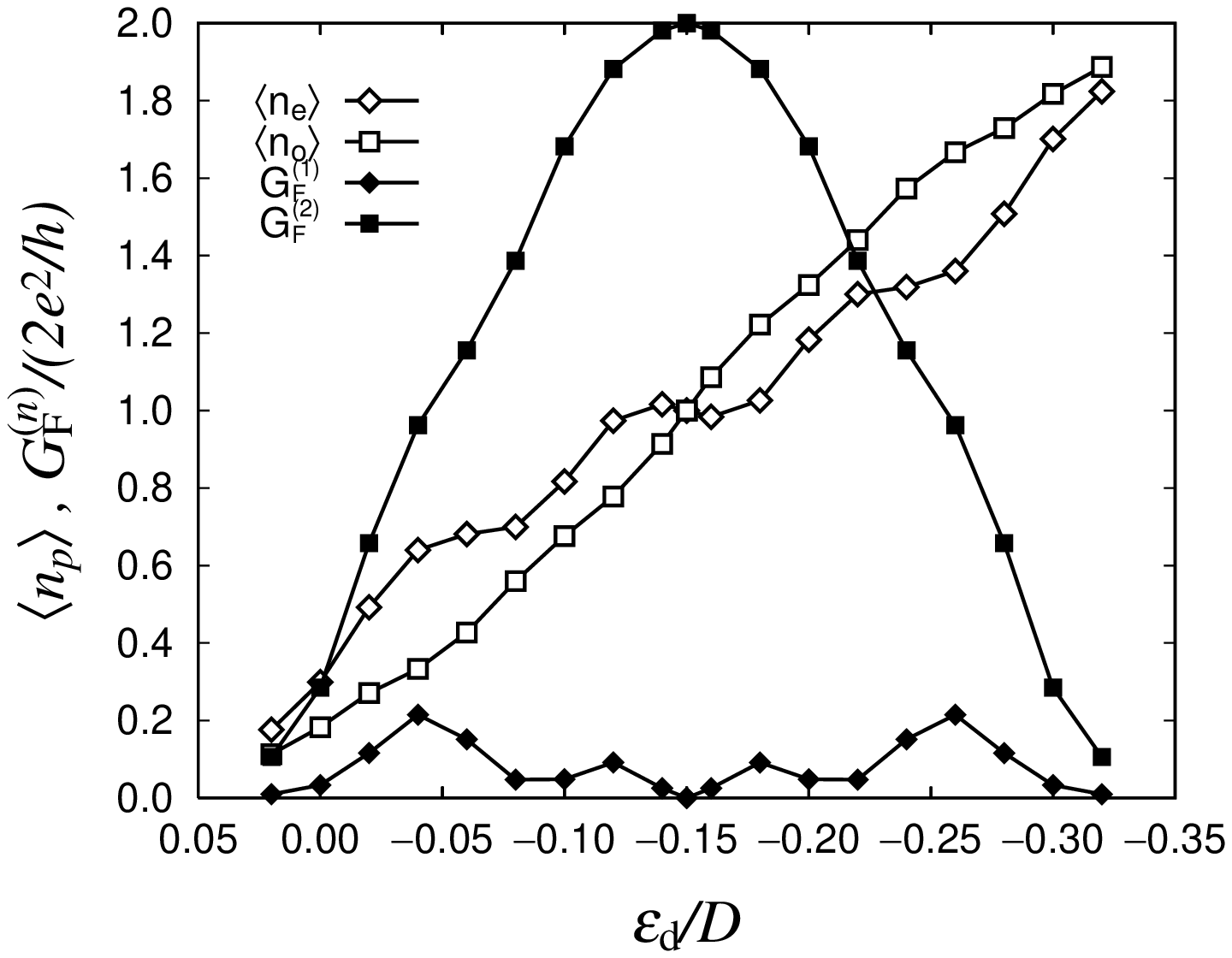}}
   \caption
   {
    The expectation value of the electron occupation numbers $\langle n_{p} \rangle$
    for the even ($p={\rm e}$) and the odd ($p={\rm o}$) orbitals,
    and the conductance $G_{\rm F}^{(n)}$
    for the THi case ($n=1$) and the THii case ($n=2$),
    as a function of the potential $\varepsilon_{\rm d}$
    at $T=0$ for the degenerate orbital case.
   }
   \label{fig:exn_2lv_od}
\end{figure}
Small but finite difference of the occupation numbers
$\langle n_{\rm e} \rangle - \langle n_{\rm o} \rangle$ is
caused by the difference between $\Delta_{\rm e}$ and
$\Delta_{\rm o}$.
The two regions, $\langle n_{\rm e} \rangle > \langle n_{\rm
o} \rangle$ and $\langle n_{\rm e} \rangle < \langle n_{\rm
o} \rangle$, are separated at $\varepsilon_{\rm d} = -0.15$
because the effective energy levels of the even and the odd
orbitals coincide at this point.
Conductance at zero temperature, $G_{\rm F}^{(1)}$, is also
presented in Fig. {\ref{fig:exn_2lv_od}}, and is small in
all region, especially near the potential $\varepsilon_{\rm
d} \sim -0.15$ since $\langle n_{\rm e} \rangle - \langle
n_{\rm o} \rangle$ is small.
We note that the factor $\langle n_{\rm e} \rangle - \langle
n_{\rm o} \rangle$ in eq. (\ref{eq:G_F}) is caused by the
interference between the tunneling processes {\it via} the
even and the odd orbitals.
The result at $T=0$ is contrasted with the non-degenerate
orbital case shown in \S \ref{sec:result_1_1}, and is also
contrasted with $G_{\rm F}^{(2)}$ for the case of THii which
will be discussed in \S \ref{sec:Two_channel}.

Next, we present the temperature dependence of the
conductance in Fig. {\ref{fig:cond_2lv_od}}.
The single particle excitation spectra for the potential
$\varepsilon_{\rm d} = -0.15$ ($\langle n_{\rm e} \rangle +
\langle n_{\rm o} \rangle = 2.0$) are shown in Fig. 
{\ref{fig:spctr_ed=-0.15.od}}, and the Kondo temperature as
a function of the potential $\varepsilon_{\rm d}$ is also
shown in Fig. {\ref{fig:tk_od}}.
\begin{figure}[htb]
   \centerline{\epsfxsize=3.25in\epsfbox{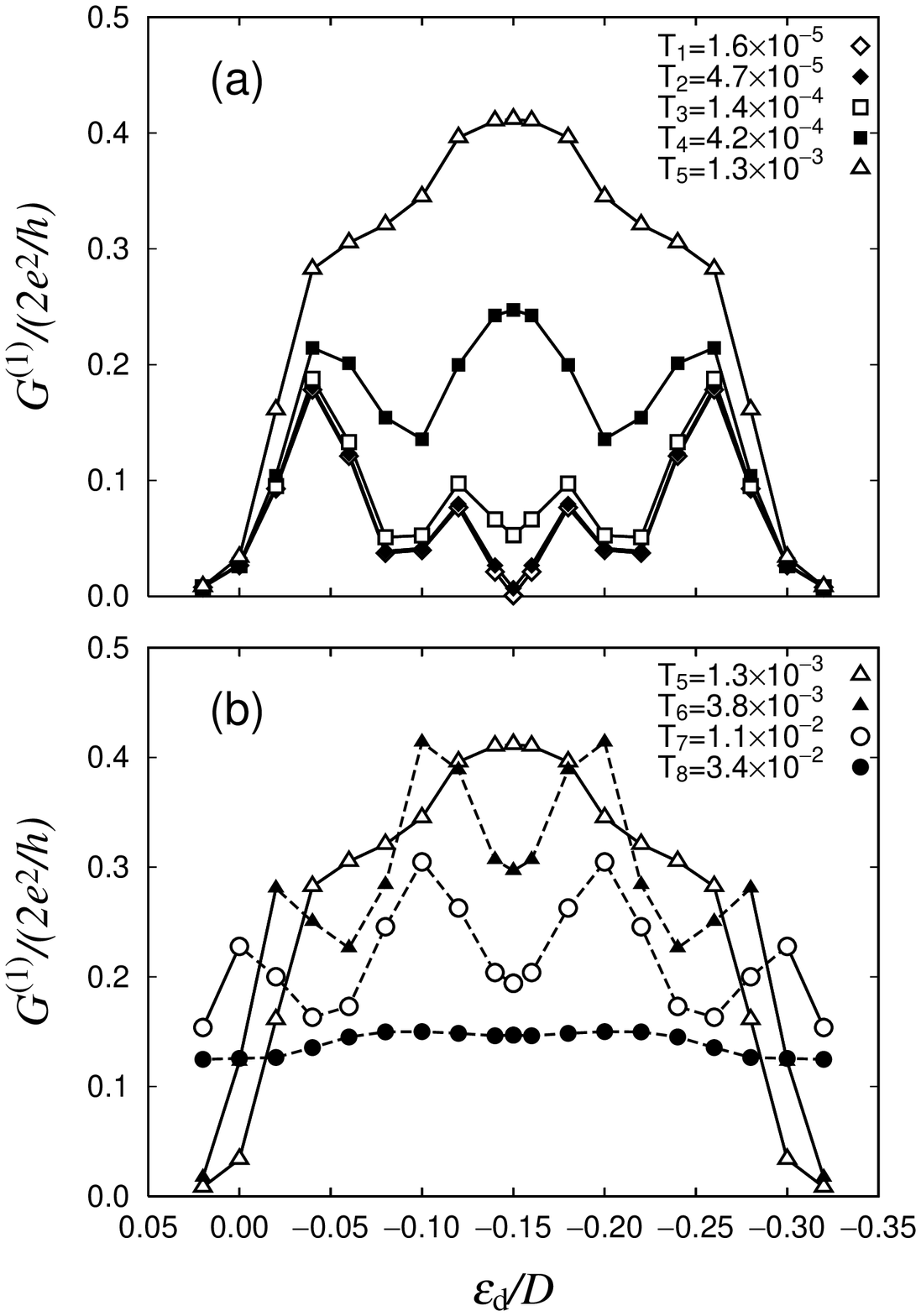}}
   \caption
   {
    The temperature dependence of the conductance
    as a function of the parameter $\varepsilon_{\rm d}$
    for the THi-2 case.
    (For the degenerate orbital case.)
    The regions given by broken lines are the magnetic state 
    and solid lines region are the non-magnetic state (spin singlet state).
    The suffixes for the temperature are common throughout \S \ref{sec:result_1_2}
    and Fig. \ref{fig:cond_2ch_od} in \S \ref{sec:Two_channel}.
   }
   \label{fig:cond_2lv_od}
\end{figure}
\begin{fullfigure}[htb]
   \centerline{\epsfxsize=7.00in\epsfbox{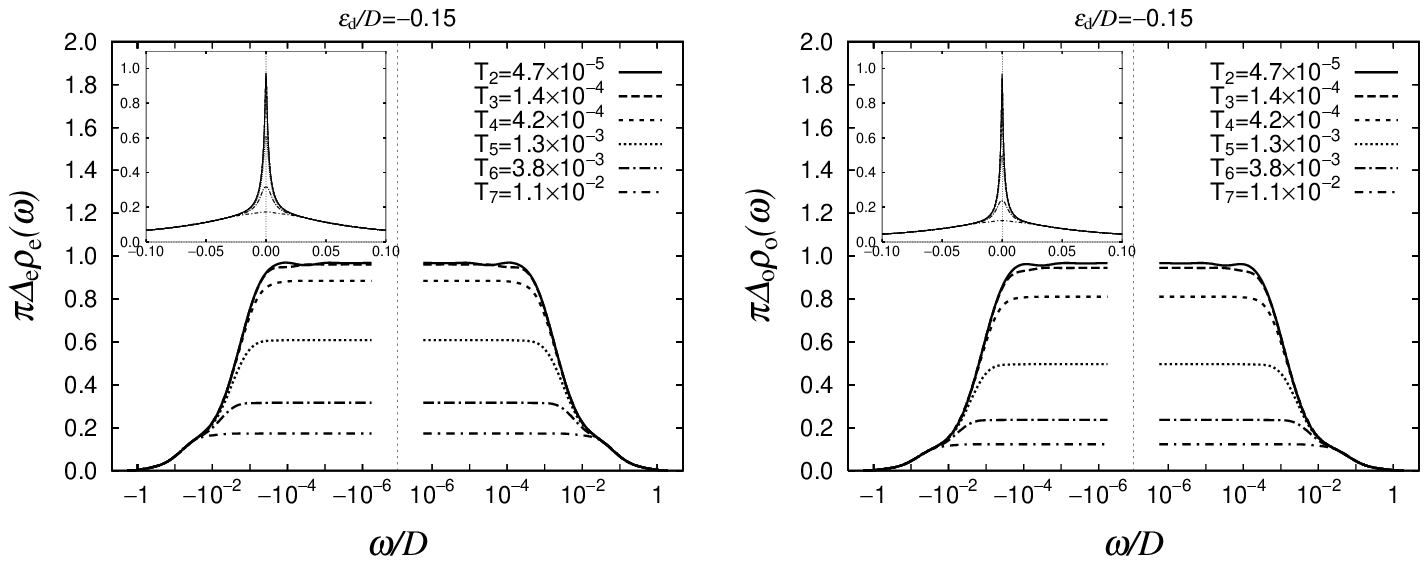}}
   \caption
   {
    The single particle excitation spectra
    at several temperatures 
    for the potential $\varepsilon_{\rm d} = -0.15$ 
    ($\langle n_{\rm e} \rangle = \langle n_{\rm o} \rangle = 1.0$).
    The results are for the degenerate orbital case.
   }
   \label{fig:spctr_ed=-0.15.od}
\end{fullfigure}
\begin{figure}[htb]
   \centerline{\epsfxsize=3.25in\epsfbox{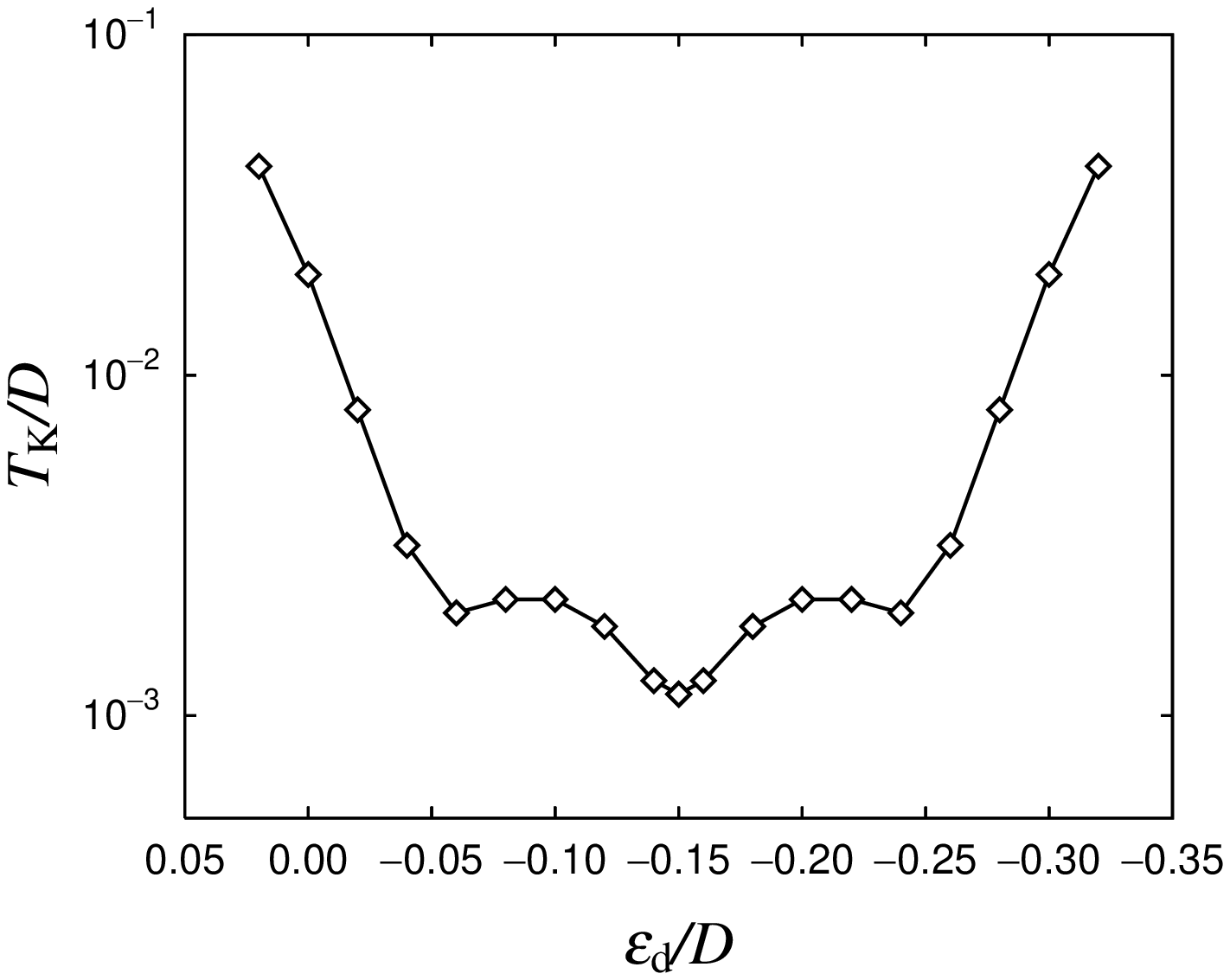}}
   \caption
   {
    The Kondo temperature as a function 
    of the potential $\varepsilon_{\rm d}$
    for the degenerate orbital case.
   }
   \label{fig:tk_od}
\end{figure}
In high temperature region (for example $T_{7} = 1.1 \times
10^{-2}$) of Fig. \ref{fig:cond_2lv_od}, there are the
Coulomb oscillations peaks near $\varepsilon_{\rm d} \sim
0.00, -0.10, -0.20$ and $-0.30$, which correspond to cross
points of energies of states with different electron
numbers; $\varepsilon_{\rm d} \sim 0, -U, -2U$ and $-3U$,
respectively.
When the temperature gradually falls down, the conductance
in the potential region $-3U \lsim \varepsilon_{\rm d} \lsim
0$ once increases, and then decreases to approach to $G_{\rm
F}^{(1)}$ at $T=0$.
This up-and-down behavior of the conductance with decreasing
temperature seems to reflect the partial growth of the
coherency in each orbital channel.
For example at $T=T_{5}$, the peak height of the Kondo
resonance in even channel is about $60 \%$ of that of the
$T=0$ limit while $50 \%$ in odd channel as seen from Fig. 
{\ref{fig:spctr_ed=-0.15.od}}.
We note the relation, $T_{5}=1.3 \times 10^{-3} \sim T_{\rm
K}^{*} \equiv T_{\rm K}(\varepsilon_{\rm d}=-0.15)=1.2
\times 10^{-3}$.
When the temperature decreases below the Kondo temperature,
the interference cancellation between the even and the odd
orbital channels becomes gradually complete, and the
conductance decreases to approach to $G_{\rm F}^{(1)}$.
At about $T \lsim T_{3} = 1.4 \times 10^{-4} \sim 0.1 T_{\rm
K}^{*}$, the conductance reaches to the low temperature
limit.
We have weak four peak structures in $G_{\rm F}^{(1)}$, but
the peak positions do not coincide with those of the Coulomb
oscillations.

We note that the minimum of the Kondo temperature, $T_{\rm
K}^{*}$, is much higher than that of the non-degenerate
orbital case presented in the previous subsection, even
though the hybridization strengths are same.
This behavior has been recognized in the 4-fold degeneracy
cases of the impurity Anderson model where the parameters
satisfy $\Delta_{\rm e} = \Delta_{\rm
o}$.~\cite{rf:Rev.Kondo_Effect,rf:Bickers}

\subsubsection{Effect of Hund's rule coupling}
\label{sec:result_1_3}

The energy splitting due to Hund's rule coupling has been
observed in the quantum dot which have high geometrical
symmetry.~\cite{rf:Tarucha}
This coupling brings the triplet spin state for the two
electrons configuration of the dot.
The Kondo temperature would be strongly reduced because the
hybridization processes are restricted in this
case.~\cite{rf:OkadaYosida}
In this subsection the numerical results for the presence of
Hund's rule coupling are shown.
We choose the parameters $\Delta_{\rm e} = 0.003 \pi$,
$\Delta_{\rm o} = 0.002 \pi$, $\Delta\varepsilon_{\rm d}=0$,
$U=0.1$ and $J_{\rm H}=0.020$.

The numerical results of $\langle n_{p} \rangle$ and $G_{\rm
F}^{(1)}$ as a function of the potential $\varepsilon_{\rm
d}$ are shown in Fig. {\ref{fig:exn_2lv_Hund}}.
\begin{figure}[htb]
   \centerline{\epsfxsize=3.25in\epsfbox{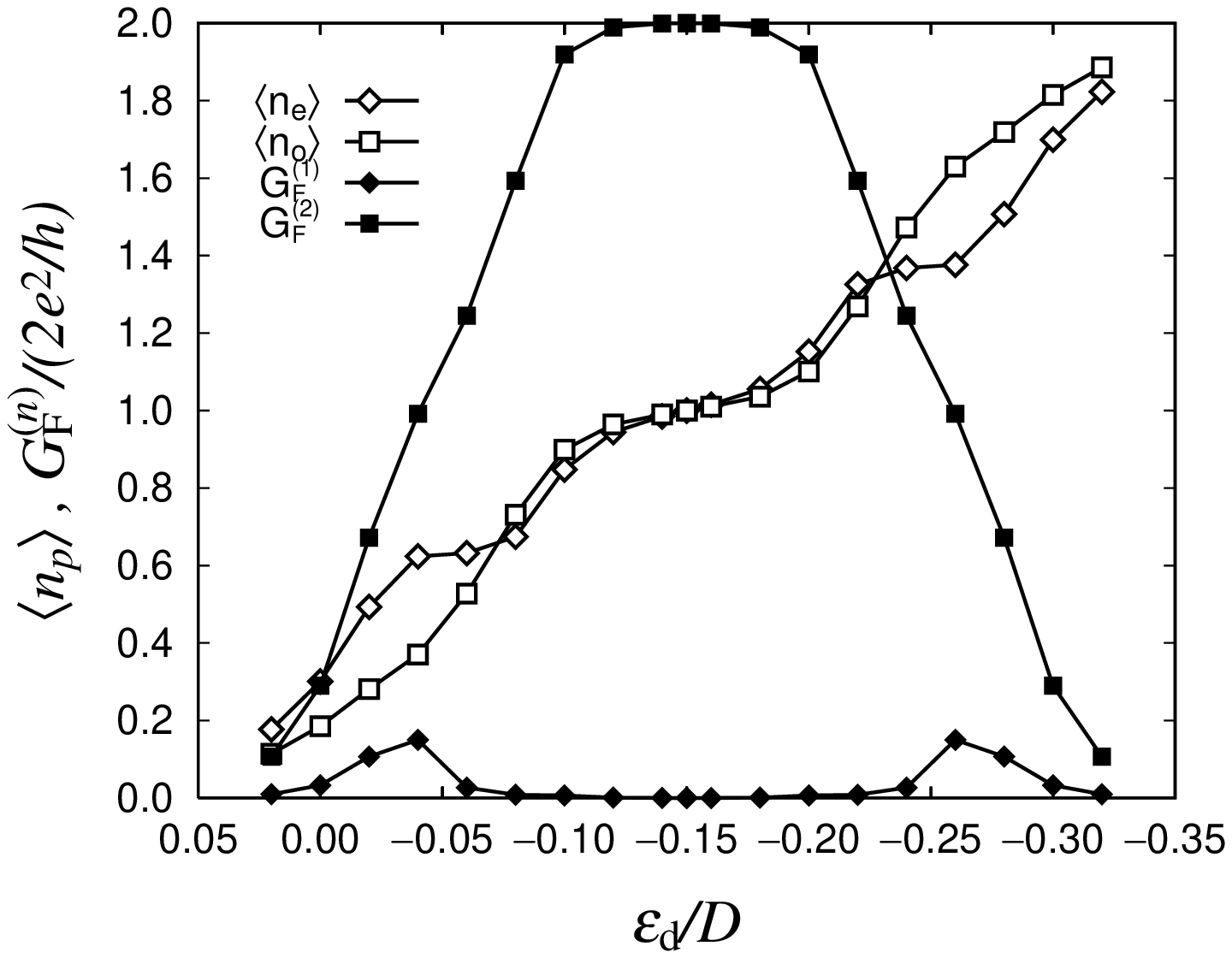}}
   \caption
   {
    The expectation value of the electron occupation numbers $\langle n_{p} \rangle$
    for the even ($p={\rm e}$) and the odd ($p={\rm o}$) orbital,
    and the conductance $G_{\rm F}^{(n)}$
    for the THi case ($n=1$) and the THii case ($n=2$),
    as a function of 
    the potential $\varepsilon_{\rm d}$ 
    at $T=0$ for the presence of Hund's rule coupling case.
   }
   \label{fig:exn_2lv_Hund}
\end{figure}
When the potential $\varepsilon_{\rm d}$ is dropped, the
first electron occupies the dot at $\varepsilon_{\rm d} \sim
0.00$ ($\langle n_{\rm e} \rangle + \langle n_{\rm o}
\rangle \sim 0.5$), the second electron at $\varepsilon_{\rm
d} \sim -U + \left| J_{\rm H} \right| = -0.08$ ($\langle
n_{\rm e} \rangle + \langle n_{\rm o} \rangle \sim 1.5$),
the third electron at $\varepsilon_{\rm d} \sim -2U - \left|
J_{\rm H} \right| = -0.22$ ($\langle n_{\rm e} \rangle +
\langle n_{\rm o} \rangle \sim 2.5$), and the forth electron
at $\varepsilon_{\rm d} \sim -3U = -0.30$ ($\langle n_{\rm
e} \rangle + \langle n_{\rm o} \rangle \sim 3.5$).
(The potential values at $\varepsilon_{\rm d} = 0, -U +
\left| J_{\rm H} \right|, - 2U - \left| J_{\rm H} \right|$
and $-3U$ are the cross points of the energies of the states
with different electron number without the hybridization
term.)
The difference of the occupation numbers $\langle n_{\rm e}
\rangle - \langle n_{\rm o} \rangle$ is brought from the
difference of the hybridization strengths as previously
noted.
The way of the electrons occupancy and the behavior of the
conductance as a function of the potential $\varepsilon_{\rm
d}$ at zero temperature, $G_{\rm F}^{(1)}$, are roughly
similar to that of the case shown in \S
\ref{sec:result_1_2}, in which Hund's rule coupling energy
is neglected.
But the effect of Hund's rule coupling can be recognized in
the potential region $-2U - \left| J_{\rm H} \right| \lsim
\varepsilon_{\rm d} \lsim -U + \left| J_{\rm H} \right|$
(where $\langle n_{\rm e} \rangle + \langle n_{\rm o}
\rangle \sim 2.0$).
The difference of the occupation number between the even and
the odd orbitals $\langle n_{\rm e} \rangle - \langle n_{\rm
o} \rangle$ is suppressed compared with the results in \S
\ref{sec:result_1_2}.
Therefore the conductance at zero temperature, $G_{\rm
F}^{(1)}$, is very small in the potential region where the
interference cancellation occurs.

The temperature dependence of the conductance is presented
in Fig. {\ref{fig:cond_2lv_Hund}}.
\begin{figure}[htb]
   \centerline{\epsfxsize=3.25in\epsfbox{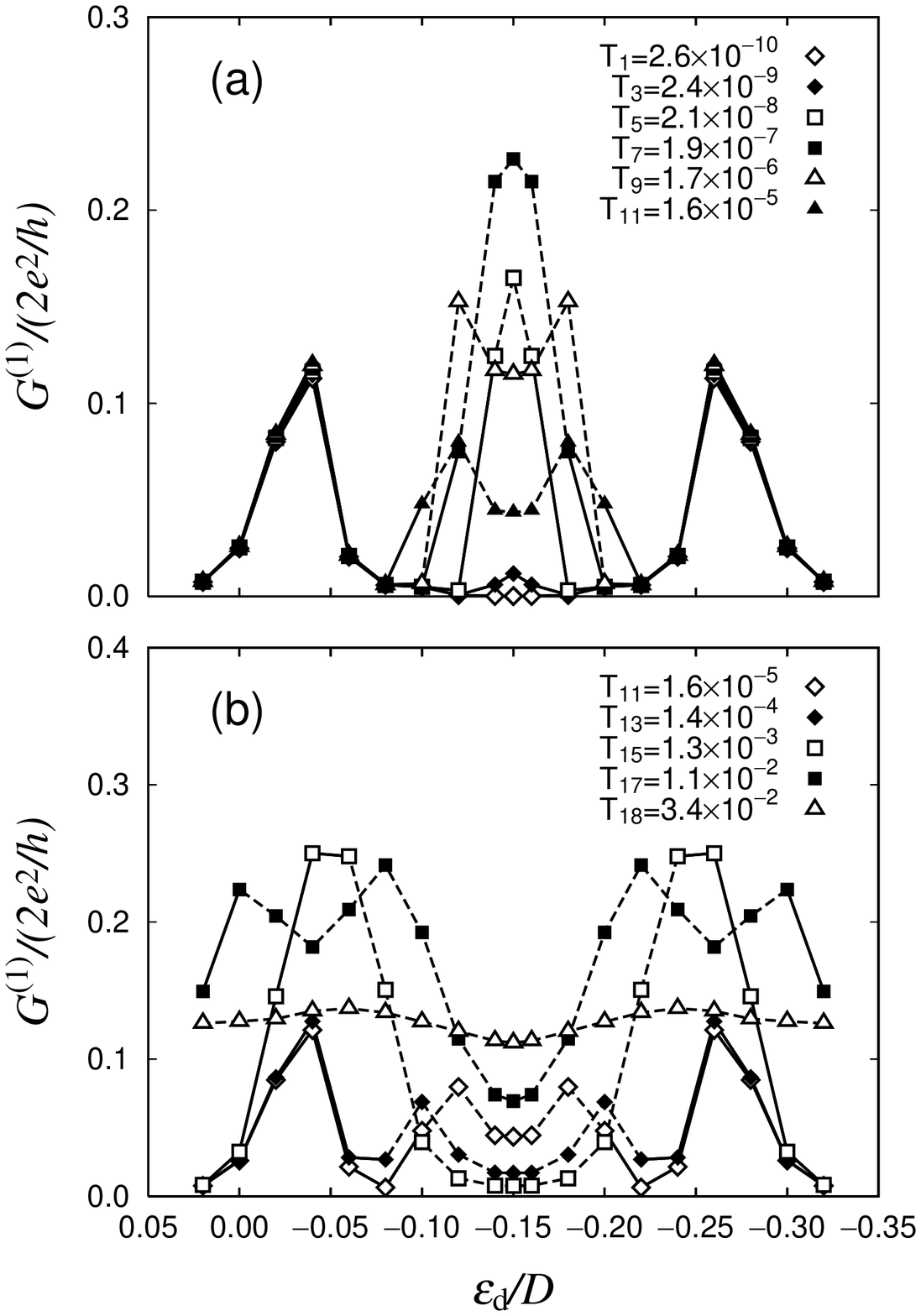}}
   \caption
   {
    The temperature dependence of the conductance
    as a function of the parameter $\varepsilon_{\rm d}$
    for the THii-3 case.
    (Hund's rule coupling case.)
    The regions given by broken lines are the magnetic state 
    and solid lines region are the non-magnetic state (spin singlet state).
    The suffixes for the temperature are common throughout \S \ref{sec:result_1_3}
    and Fig. \ref{fig:cond_2ch_Hund} in \S \ref{sec:Two_channel}.
   }
   \label{fig:cond_2lv_Hund}
\end{figure}
The single particle excitation spectra at the potential
$\varepsilon_{\rm d} = -0.15$ ($\langle n_{\rm e} \rangle +
\langle n_{\rm o} \rangle = 2.0$) are shown in Fig. 
{\ref{fig:spctr_ed=-0.15.Hund}}, and the Kondo temperature
as a function of the potential $\varepsilon_{\rm d}$ is
shown in Fig. {\ref{fig:tk_Hund}}.
\begin{fullfigure}[htb]
   \centerline{\epsfxsize=7.00in\epsfbox{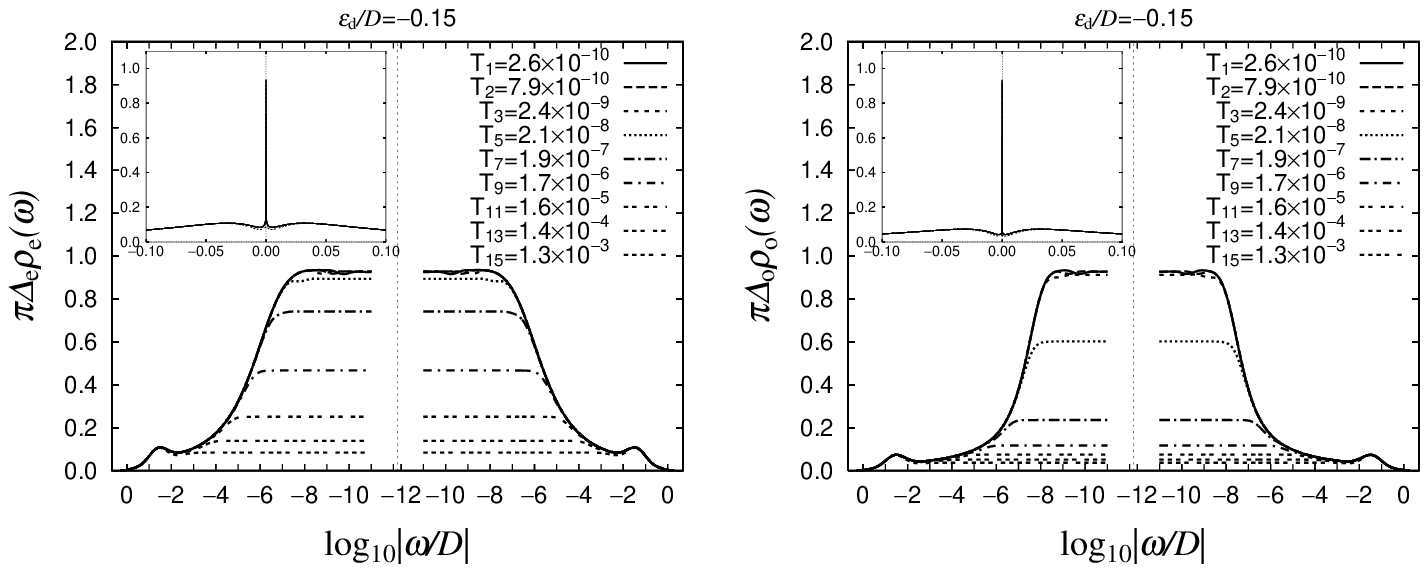}}
   \caption
   {
    The single particle excitation spectra at several temperatures 
    for the potential $\varepsilon_{\rm d} = -0.15$
    ($\langle n_{\rm e} \rangle + \langle n_{\rm o} \rangle =2.0$).
    The results are for the presence of Hund's rule coupling case.
   }
   \label{fig:spctr_ed=-0.15.Hund}
\end{fullfigure}
\begin{figure}[htb]
   \centerline{\epsfxsize=3.25in\epsfbox{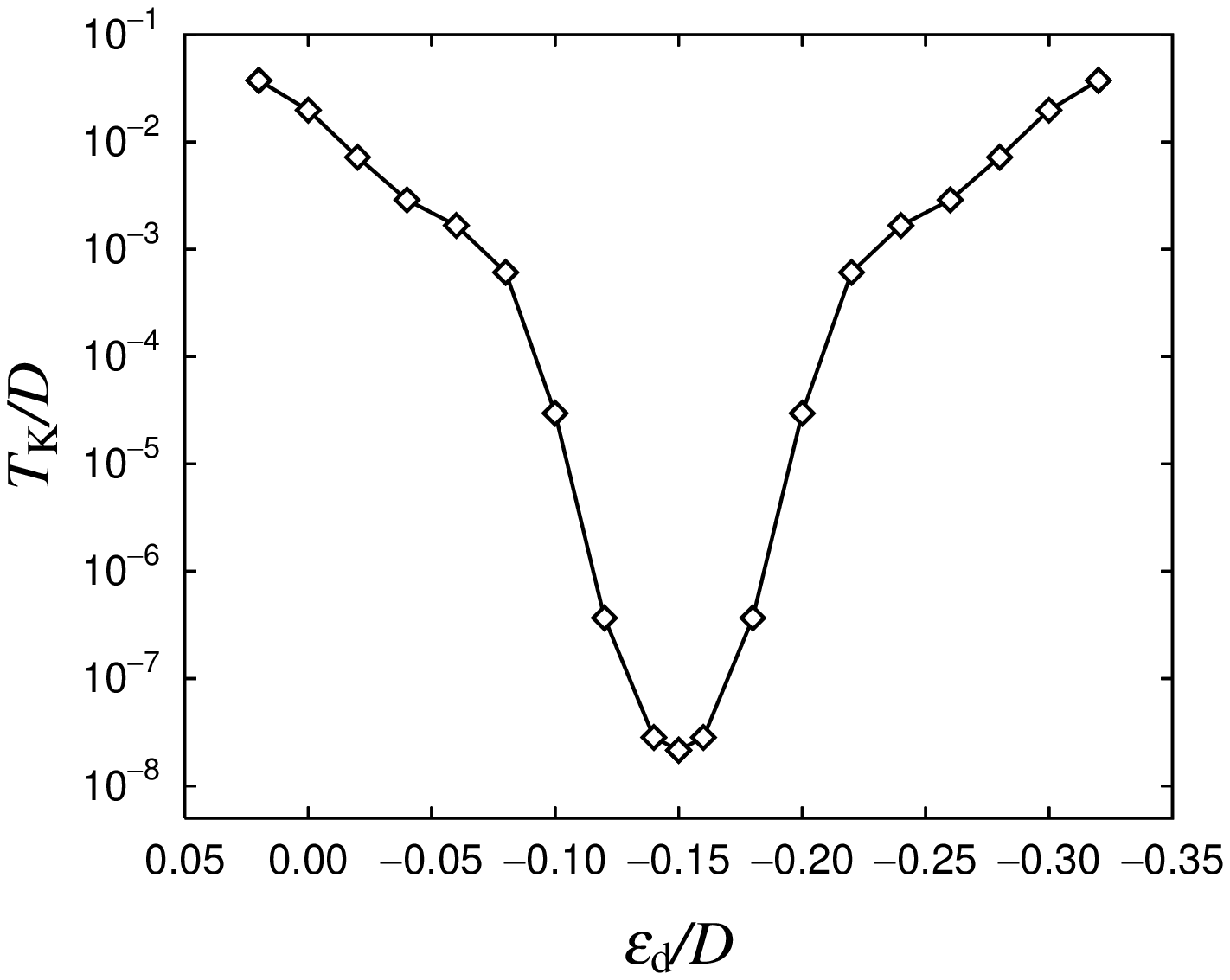}}
   \caption
   {
    The Kondo temperature as a function of the potential $\varepsilon_{\rm d}$
    for the presence of Hund's rule coupling case.
   }
   \label{fig:tk_Hund}
\end{figure}

The Kondo temperature in the region $ -2U - \left| J_{\rm H}
\right| \lsim \varepsilon_{\rm d} \lsim -U + \left| J_{\rm
H} \right|$ is far less than that shown in Fig. 
\ref{fig:tk_od}.
This small $T_{\rm K}$ is caused by the reduction of the
hybridization processes due to Hund's rule coupling which
restricts the two electrons states in the dot to be the
triplet state.

In high temperature region (for example $T_{17} = 1.1 \times
10^{-2}$) of Fig. \ref{fig:cond_2lv_Hund}, there are the
Coulomb oscillations at $\varepsilon_{\rm d} \sim 0.00,
-0.08, -0.22$ and $\varepsilon_{\rm d} \sim -0.30$.
The conductance approaches to $G_{\rm F}^{(1)}$ at about $T
\sim T_{3} \sim 0.1 T_{\rm K}^{*}$ ($T_{\rm K}^{*} \equiv
T_{\rm K}(\varepsilon_{\rm d} = -0.15) = 2.1 \times
10^{-8}$).

We compare the results in Fig. {\ref{fig:cond_2lv_Hund}}
with those in Fig. {\ref{fig:cond_2lv_od}}.
The conductance in the two cases shows almost identical
behavior in the region $\varepsilon_{\rm d} \gsim -U +
\left| J_{\rm H} \right| = -0.08$ and $\varepsilon_{\rm d}
\lsim -2U - \left| J_{\rm H} \right| = -0.22$.
They reach to the lowest temperature behavior for $T \lsim
10^{-4}$ in that region.
In the region $-2U - \left| J_{\rm H} \right| \lsim
\varepsilon_{\rm d} \lsim -U + \left| J_{\rm H} \right|$ in
Fig. {\ref{fig:cond_2lv_Hund}}, the conductance initially
increases with decreasing temperature, and then it decreases
in the very low temperature.
This behavior is similar to the up-and-down behavior in the
region of $-2U \lsim \varepsilon_{\rm d} \lsim -U$ in Fig. 
{\ref{fig:cond_2lv_od}}.
However in the present case, the temperature at which the
conductance reaches to the lowest temperature behavior is
much lower because $T_{\rm K}^{*}$ is very low.
The difference of the occupation number is suppressed by
Hund's rule coupling, so the conductance $G_{\rm F}^{(1)}$
in this region becomes almost zero.

The qualitative features of the temperature dependence of
the spectra shown in Fig. \ref{fig:spctr_ed=-0.15.Hund} are
similar to those shown in Fig. \ref{fig:spctr_ed=-0.15.od}
except that the temperature scale of the former is very low.
The peak at $\omega \sim 0$ in $\rho_{\rm e}(\omega)$
gradually begins to grow up at about $T \sim 10^{-5}$, while
the peak in $\rho_{\rm o}(\omega)$ starts to grow up at
about $T \sim 10^{-7}$.
The difference of the temperatures at which the Kondo peak
starts to grow up in each channel is enhanced by Hund's rule
term, though $\Delta_{\rm e}$ and $\Delta_{\rm o}$ are
chosen commonly to those of Fig. 
\ref{fig:spctr_ed=-0.15.od}.
But we note that the peaks reach almost simultaneously to
the low temperature limit at about $T \sim 10^{-9}$.
The finite conductance value in the temperature region
$10^{-7} \lsim T \lsim 10^{-5}$ reflects the growth of the
peak in the even channel.
The decrease of the conductance in $10^{-9} \lsim T \lsim
10^{-8}$ reflects the growth of the peak in the odd channel.
Hund's rule coupling prevents to observe the low temperature
limit behaviors, but it makes wide an intermediate
temperature region where a part of the orbital is affected
by the Kondo effect.

\section{Two Conduction Channels in Each Lead}
\label{sec:Two_channel}

In \S \ref{sec:One_channel}, we have discussed the situation
classified to the THi that dot's orbitals hybridize only to
one-dimensional degree of freedom in each lead.
In this section we consider also the perpendicular
components in the leads as the degree of freedoms.
We assume that there are two orbitals in the dot, which are
even and odd, respectively, under the $zx$ mirror.
These are assumed to be even under the $yz$ mirror.
We name these orbitals in the dot as the even orbital and
the odd orbital, and denote them by $d_{{\rm e} \sigma}$ and
$d_{{\rm o} \sigma}$.
(Note that in \S \ref{sec:One_channel}, the even and the odd
were denoted with respect to the $yz$ plane.)
We assume that we have two conduction channels in each lead,
which are even and odd, respectively, under the $zx$ mirror.
These are named as the even and the odd channels in the left
and the right leads.
We write them as $c_{{\rm L e} k \sigma}$, $c_{{\rm L o} k
\sigma}$, $c_{{\rm R e} k \sigma}$ and $c_{{\rm R o} k
\sigma}$, respectively.
The Hamiltonian $H_{\rm l}$ and $H_{\rm l-d}$ is written as
follows;
\begin{eqnarray}
   H_{{\rm l}} & = & 
                     \sum_{k \sigma}(
                     \varepsilon_{k}
                     c_{{\rm L e} k \sigma}^{\dagger}
                     c_{{\rm L e} k \sigma}
                 +
                     \varepsilon_{k}
                     c_{{\rm L o} k \sigma}^{\dagger}
                     c_{{\rm L o} k \sigma}) 
                 \nonumber\\
               &&   +
                     \sum_{k \sigma}(
                     \varepsilon_{k}
                     c_{{\rm R e} k \sigma}^{\dagger}
                     c_{{\rm R e} k \sigma},
                 +
                     \varepsilon_{k}
                     c_{{\rm R o} k \sigma}^{\dagger}
                     c_{{\rm R o} k \sigma}),
   \label{eq:H_l_2}\\
   H_{{\rm l-d}} & = & 
                       \frac{1}{\sqrt{2}}
                       \sum_{k \sigma}
                       (V_{\rm e} 
                       d_{{\rm e} \sigma}^{\dagger} c_{{\rm L e} k \sigma}
                   +
                        V_{\rm o}
                       d_{{\rm o} \sigma}^{\dagger} c_{{\rm L o} k \sigma}
                       + {\rm h.c.}) \nonumber \\
                   && 
                   +
                       \frac{1}{\sqrt{2}}
                       \sum_{k \sigma}
                       (V_{\rm e}
                       d_{{\rm e} \sigma}^{\dagger} c_{{\rm R e} k \sigma}
                   +    
                        V_{\rm o}
                       d_{{\rm o} \sigma}^{\dagger} c_{{\rm R o} k \sigma}
                       + {\rm h.c.}).
   \label{eq:H_l-d_2}
\end{eqnarray}
When the operators, $s_{k \sigma}$ and $a_{k \sigma}$ in
eqs. (\ref{eq:H_l_1}) and (\ref{eq:H_l-d_1}), are replaced
respectively by newly defined ones, $s_{k \sigma} = (c_{{\rm
L e} k \sigma} + c_{{\rm R e} k \sigma}) / \sqrt{2}$ and
$a_{k \sigma} = (c_{{\rm L o} k \sigma} + c_{{\rm R o} k
\sigma}) / \sqrt{2}$, we have the identical form for the
Hamiltonian.

But the conductance formula is different from that in \S
\ref{sec:One_channel}.
We have the situation classified to the THii that the
tunneling {\it via} the even and the odd channels are
independent in the tunneling between leads.
The conductance formula is given as
follows,~\cite{rf:Oguri.2}
\begin{eqnarray}
   G^{(2)} & = & \frac{2e^{2}}{h} 
                  \sum_{p={\rm e, o}}
                  \int d\varepsilon 
                       \left( -\frac{\partial f}{\partial \varepsilon} \right)
                       \pi \Delta_{p} \rho_{p}(\varepsilon).
   \label{eq:G2}
\end{eqnarray}
Here $\rho_{p}(\varepsilon)$ is the single particle
excitation spectrum of the dot orbital with symmetry $p$.
At $T=0$, the conductance formula is reduced to the
following expression,
\begin{eqnarray}
   G_{\rm F}^{(2)} & = & \frac{2e^{2}}{h} 
                  \sum_{p={\rm e, o}}
                  \sin^{2} 
                       \left\{ \frac{\pi}{2} 
                               \langle n_{p} \rangle
                       \right\}.
   \label{eq:G2_F}
\end{eqnarray}

The numerical calculations are performed for the three
cases; (1) non-degenerate orbital case, (2) degenerate
orbital case, and (3) presence of Hund's rule coupling.
The parameters are chosen to be same as in \S
\ref{sec:One_channel}, for comparison.
Then the electronic states of the both systems are
identically mapped.
The conductance for three cases is shown in Figs. 
{\ref{fig:cond_2ch_nod}}, {\ref{fig:cond_2ch_od}} and
{\ref{fig:cond_2ch_Hund}}, respectively.
(Conductance at $T=0$ has been shown in Figs. 
\ref{fig:exn_2lv_nod},
\ref{fig:exn_2lv_od}
and
\ref{fig:exn_2lv_Hund},
respectively.)
\begin{figure}[htb]
   \centerline{\epsfxsize=3.25in\epsfbox{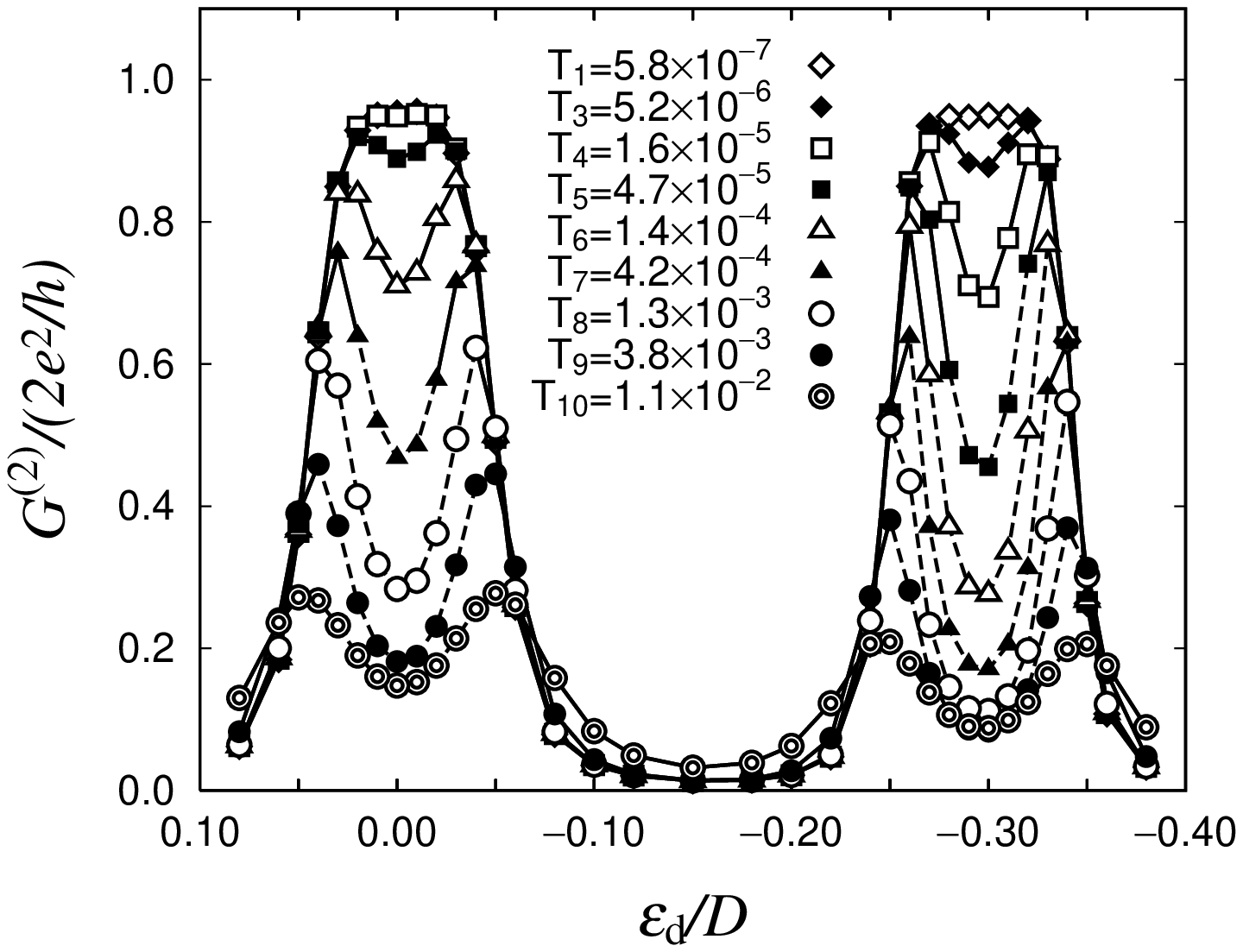}}
   \caption
   {
    The temperature dependence of the conductance
    as a function of the parameter $\varepsilon_{\rm d}$
    for the THii-1 case.
    (Tunneling processes {\it via} the even and the odd orbitals are independent.
    For the non-degenerate orbital case.)
   }
   \label{fig:cond_2ch_nod}
\end{figure}
\begin{figure}[htb]
   \centerline{\epsfxsize=3.25in\epsfbox{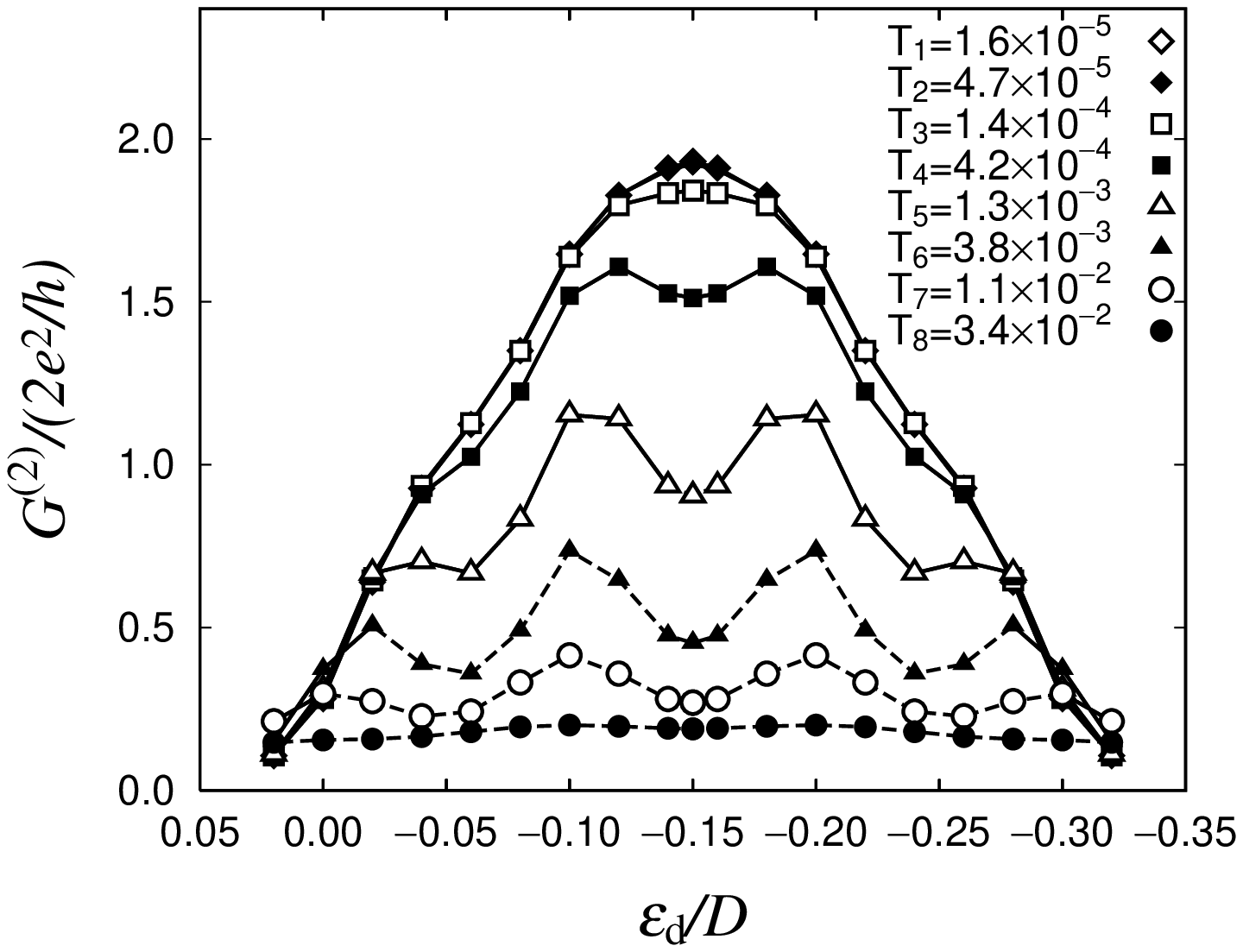}}
   \caption
   {
    The temperature dependence of the conductance
    as a function of the parameter $\varepsilon_{\rm d}$
    for the THii-2 case.
    (For the degenerate orbital case.)
   }
   \label{fig:cond_2ch_od}
\end{figure}
\begin{figure}[htb]
   \centerline{\epsfxsize=3.25in\epsfbox{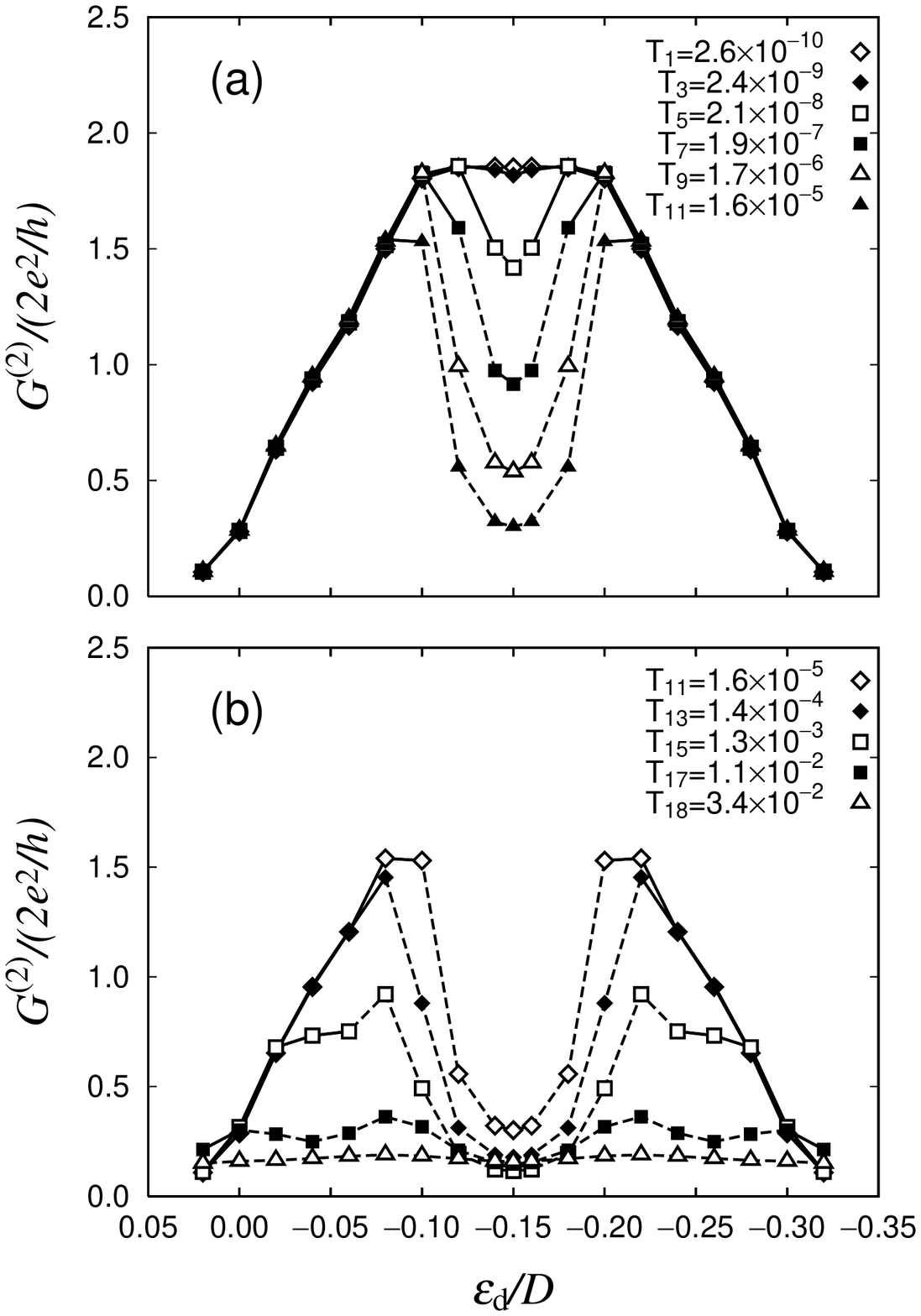}}
   \caption
   {
    The temperature dependence of the conductance
    as a function of the parameter $\varepsilon_{\rm d}$
    for the THii-3 case.
    (Hund's rule coupling case.)
   }
   \label{fig:cond_2ch_Hund}
\end{figure}
As note previously, the interference effect of tunneling
processes {\it via} the even and odd orbitals do not occur
in the present THii case.
Therefore the phenomena due to the interference in THi case
do not appear in the present.
Maximum of the intensity at $T=0$ has $2 \times 2e^{2} /h$,
twice of the unitarity limit value of the single conduction
channel case.
In Fig. {\ref{fig:cond_2ch_nod}}, the differences of the
conductance behaviors from those of Fig. 
\ref{fig:cond_2lv_nod} are not well recognized, because the
interference is not so significant in the non-degenerate
orbital case.

In Fig. {\ref{fig:cond_2ch_od}} of the degenerate orbital
case, the conductance in high temperature region (for
example $T_{7} = 1.1 \times 10^{-2}$) behaves the Coulomb
oscillations type similar to that of Fig. 
\ref{fig:cond_2lv_od}.
As temperature decreases, it increases gradually filling the
valley between four peaks.
At $T \sim T_{5} = 1.3 \times 10^{-3}$, the conductance is
rather large indicating the partial growing up of the Kondo
resonance in each channel.
But we do not have the up-and-down behaviors such that shown
in Fig. \ref{fig:cond_2lv_od}.
At about $T \lsim T_{3} = 1.4 \times 10^{-4} \sim 0.1 T_{\rm
K}^{*}$, the conductance approaches to the low temperature
limit which has a broadened peak structure.
In Fig. {\ref{fig:cond_2ch_Hund}}, for Hund's rule coupling
case, the conductance behavior is similar to that of the
degenerate orbital case.
But the conductance in the potential region
$\varepsilon_{\rm d} \sim -0.15$ does not increases to the
low temperature limit values until the temperature becomes
very low, $T \lsim T_{3} = 2.4 \times 10^{-9} \sim 0.1
T_{\rm K}^{*}$.
Therefore in the wide temperature range we can see two peak
structures similar to that of the non-degenerate orbital
case.

\section{Summary and Discussion}
\label{sec:Summary}

The Kondo effects in the quantum dot systems have been
investigated for the cases that (THi) there is one
conduction channel in each lead and the tunneling processes
{\it via} the different orbitals in the dot interfere each
other (in \S \ref{sec:One_channel}), and (THii) there are
two conduction channels in each lead but the tunneling
processes do not interfere (in \S \ref{sec:Two_channel}).
In both cases, the following three cases for the orbital
degeneracy have been examined; (1) non-degenerate orbital
case, (2) degenerate orbital case, and (3) presence of
Hund's rule coupling case.
The linear response conductance for the bias voltage has
been calculated as a function of the gate voltage.
The numerical calculations have been performed in wide
temperature range by using NRG method.

The Coulomb oscillations are observed in high temperature
region commonly in the all cases, but various types of
conductance behavior appear reflecting the growing up of the
Kondo resonance at low temperatures.
The low temperature limit behaviors appear in the
temperature region $T \lsim 0.1 T_{\rm K}^{*} \sim 0.2
T_{\rm K}^{*}$, where $T_{\rm K}^{*}$ is the minimum value
of the Kondo temperature when $\varepsilon_{\rm d}$ is
changed.

In the non-degenerate orbital cases, THi-1 and THii-1, the
pair of the two peaks of the Coulomb oscillations gradually
changes one flat peak structure as temperature decreases
below $0.2 T_{\rm K}^{*}$.
The conductance behaviors in these two cases are not so
different because the interference effect is not
significant.

In the degenerate orbital case of THi-2, the interference
cancellation between the tunneling processes {\it via} the
different orbitals causes small conductance at temperatures
below $0.1 T_{\rm K}^{*}$.
But we have the characteristic up-and-down behavior of the
conductance with increasing temperature in the intermediate
temperature range.
This is caused by the partial destruction of the coherency
due to the partial breaking of the Kondo singlet state in
each orbital channel.
In the case of THi-3 (Hund's rule coupling), the temperature
region where the conductance shows the characteristic
behaviors due to such partial destruction of the coherency
shifts to low temperature side, because the Kondo
temperature is largely reduced.

In Fig. \ref{fig:cond_2ch_od} of THii-2 and Fig. 
\ref{fig:cond_2ch_Hund} of THii-3 in \S
\ref{sec:Two_channel}, four Coulomb oscillations peaks
gradually become one broad peak structure as temperature
decreases below $0.1 T_{\rm K}^{*}$, because the
interference between two channels does not exist.
But we have a regime indicating the partial growing up of
the Kondo resonance in the intermediate temperature region.
In the case of THii-3, we have two peak structures similar
to that of the non-degenerate orbital case in wide
temperature range.

Electronic states of the cases THi and THii are mapped on
with each other, but the tunneling current in low
temperatures shows quite different behaviors.
The sensitivities of the conductance to the temperature and
to the gate voltage are caused by the difference of the
Kondo temperature at each potential, are inherent in the
interacting systems.

Here we roughly estimate the characteristic temperature by
assuming $U \sim 1 {\rm meV} (\sim 10 {\rm K})$ and
$\Delta_{p} \sim 0.1 {\rm meV} (\sim 1 {\rm K})$.
The Kondo temperature $T_{\rm K}^{*}$ is estimated to be
about $3 {\rm mK}$ ($\Delta_{\rm o} \sim 0.06 {\rm meV}$)
and $30 {\rm mK}$ ($\Delta_{\rm e} \sim 0.1 {\rm meV}$) for
the case (1), $100 {\rm mK}$ for the case (2), and $10^{-3}
{\rm mK}$ for the case (3).
We note, however, the Kondo temperature is very sensitive to
the hybridization strength.

Recently experimental work has been performed with
controlling $\Delta$ by Goldhaber-Gordon {\it et
al.}~\cite{rf:Kondo-Kastner}
The change of the conductance due to the Kondo effect seems
to appear in their system.
The results seem to correspond to our numerical results at
higher temperature region shown in Fig. 
{\ref{fig:cond_2lv_nod}} and Fig. {\ref{fig:cond_2ch_nod}},
$10^{-3} \lsim T \lsim 10^{-2}$, qualitatively.

\section*{Acknowledgments}

The authors would like to thank A. Oguri, R. Takayama and S. 
Suzuki for valuable discussion and information.
This work is supported by Grant-in-Aid No. 06244104, No. 
09244202, and No. 09640451 from Ministry of Education,
Science and Culture of Japan.
The numerical computation was performed at the Supercomputer
Center of Institute for Solid State, the Computer Center of
Institute for Molecular Science and the Computer Center of
Tohoku University.

\appendix

\section{Non-Degenerate Orbital to Degenerate Orbital}
\label{sec:App_1}

In \S \ref{sec:One_channel} and \ref{sec:Two_channel}, we
have used models that the energy separation of the orbitals,
$\Delta \varepsilon_{\rm d}$, is extreme.
It is large (in \S \ref{sec:result_1_1} and Fig. 
\ref{fig:cond_2ch_nod} in \S \ref{sec:Two_channel}) or zero
(in \S \ref{sec:result_1_2}, \ref{sec:result_1_3} and Fig. 
\ref{fig:cond_2ch_od}, \ref{fig:cond_2ch_Hund} in \S
\ref{sec:Two_channel}).
Here we change gradually $\Delta \varepsilon_{\rm d}$ and
study when the low temperature properties changes from the
non-degenerate orbital case to the degenerate orbital case.
The parameters for the hybridization strength are chosen to
satisfy the relations $\Delta_{\rm e} = \Delta_{\rm o}
\equiv \Delta$, for simplicity.
(For the case of THi, the tunneling current between leads
disappears by this setting.
But in Appendix, we mainly concentrate on the effects of the
energy separation on the Kondo effect.)
The parameters are chosen as $\Delta = 0.0025 \pi$, $U =
0.1$ and $J_{\rm H} = 0$.
The potential depth is fixed at $\varepsilon_{\rm d}=-0.15$,
therefore the system is in the electron-hole symmetry.
(The electron occupation number in the dot satisfies
$\langle n \rangle \equiv \langle n_{\rm e} \rangle +
\langle n_{\rm o} \rangle =2$, exactly.)

The electron occupation numbers for the even and the odd
orbitals are shown as a function of $\Delta \varepsilon_{\rm
d}$ in Fig. {\ref{fig:exn_2lv_nod-od}}.
\begin{figure}[htb]
   \centerline{\epsfxsize=3.25in\epsfbox{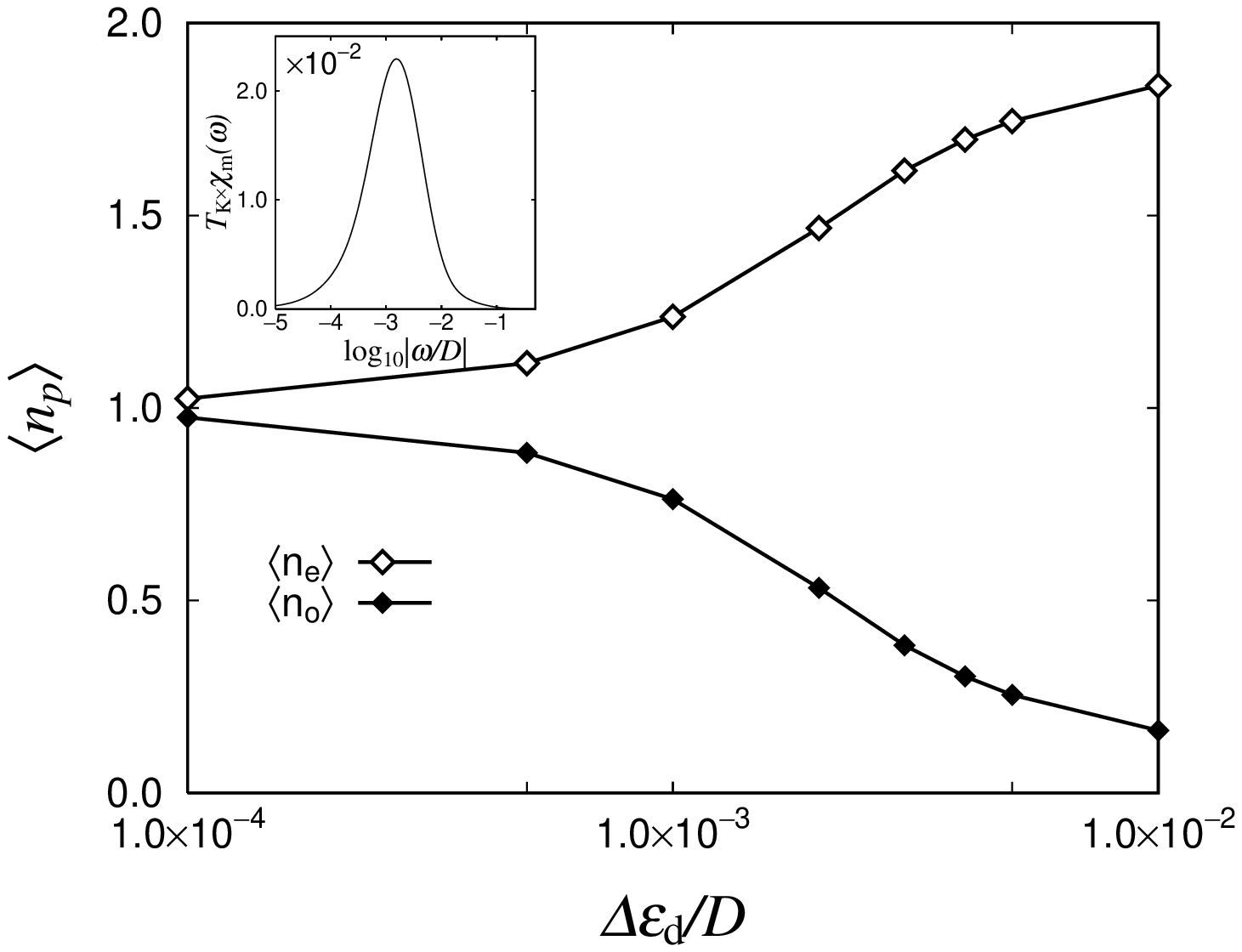}}
   \caption
   {
    The electron occupation numbers for the even and the odd orbitals
    as a function of the orbital energy separation $\Delta \varepsilon_{\rm d}$.
    Inset figure is the magnetic excitation spectrum 
    at $\varepsilon_{\rm d}=0$, $T=0$.
    The Kondo temperature at $\Delta \varepsilon_{\rm d}=0$ is estimated to be
    about $T=1.5 \times 10^{-3}$ from the peak position of the spectrum.
   }
   \label{fig:exn_2lv_nod-od}
\end{figure}
As $\Delta \varepsilon_{\rm d}$ increases, difference of the
occupation number steeply increases at about $\Delta
\varepsilon_{\rm d} \sim 2 \times 10^{-3}$ (where $\langle
n_{\rm e} \rangle \sim 1.5$ and $\langle n_{\rm o} \rangle
\sim 0.5$).
The Kondo temperature at $\Delta \varepsilon_{\rm d} = 0$ is
estimated to be about $T_{\rm K} = 1.5 \times 10^{-3}$ from
the magnetic excitation spectrum shown in the inset of Fig. 
\ref{fig:exn_2lv_nod-od}.
We may conclude that the cross over occurs at about $\Delta
\varepsilon_{\rm d} \sim T_{\rm K}(\Delta \varepsilon_{\rm
d} = 0)$.
We note that $\Delta \varepsilon_{\rm d}$ is much smaller
than the hybridization width, $\Delta$.


\begin{thebibliography}{99}
\bibitem{rf:2chlFriedel}
We note that Friedel's sum rule for each parity component is
not correct in the strict sense.
This has been already demonstrated for two impurity Anderson
model problem in ref. \citen{rf:2chlFriedel_Sakai}.
\bibitem{rf:Kondo_Temp.}
It is reasonable to use the peak position of the magnetic
excitation spectrum as the definition of the Kondo
temperature because the magnetic excitation spectrum
$\chi_{m}^{''}(\omega)$ always have the lower energy peak
than the other spectrum, as if the charge excitation
spectrum $\chi_{c}^{''}(\omega)$, even if there are no spin
moment in the local site.
\end{thebibliography}
\end{document}